\documentclass[a4paper,fleqn,usenatbib]{mnras}

\usepackage{newtxtext,newtxmath}

\usepackage[T1]{fontenc}
\usepackage{ae,aecompl}

\usepackage{graphicx}	
\usepackage{amsmath}	
\usepackage{amssymb}	
\usepackage{mathptmx}
\usepackage{mathrsfs}
\usepackage{wasysym}
\usepackage{epsfig}
\usepackage{amsmath}
\usepackage{amssymb}
\usepackage{pdflscape}
\usepackage{pifont}
\usepackage{afterpage}
\usepackage{paralist}
\usepackage{graphicx}
\usepackage{afterpage}
\usepackage{amsfonts}
\usepackage{bm}
\usepackage[flushleft]{threeparttable}
\usepackage{float}
\usepackage{array}
\newcolumntype{P}[1]{>{\centering\arraybackslash}p{#1}}
\usepackage{lscape}
\usepackage{subcaption}
\usepackage{alphalph}

\usepackage[section]{placeins}
\usepackage{url}
\usepackage{natbib}
\usepackage{multicol}
\usepackage{fancyhdr}
\let\oldhref\href
\renewcommand{\href}[2]{\oldhref{#1}{\hbox{#2}}}
\graphicspath{{SDSS+VLA+LOFAR/}}
\defcitealias{hard_lofar16}{H16}
\usepackage{color}

\title[Remnant radio-loud AGN in the Herschel-ATLAS field]{Remnant radio-loud AGN in the Herschel-ATLAS field}

\author[V. H. Mahatma et al.]{
V. H.\ Mahatma,$^{1}$\thanks{E-mail: v.mahatma2@herts.ac.uk}
M. J.\ Hardcastle,$^{1}$ 
W. L.\ Williams,$^{1}$ 
M. Brienza,$^{2,3}$
M. Br\" uggen$^{4}$,
\newauthor
J. H. Croston, $^{5}$
G. Gurkan,$^{6}$
J. J.\ Harwood,$^{2}$
M. Kunert-Bajraszewska,$^{7}$
R. Morganti,$^{2,3}$
\newauthor
H. J. A. R\" ottgering,$^{8}$	
T. W. Shimwell,$^{8}$
C. Tasse$^{9,10}$
\\
$^{1}$ Centre for Astrophysics Research, School of Physics, Astronomy and Mathematics, University of Hertfordshire, College Lane, Hatfield AL10 9AB, UK\\
$^{2}$ ASTRON, The Netherlands Institute for Radio Astronomy, Postbus 2, 7990 AA, Dwingeloo, The Netherlands\\
$^{3}$ Kapteyn Astronomical Institute, University of Groningen, P.O. Box 800, 9700 AV Groningen, The Netherlands\\
$^{4}$ Hamburger Sternwarte, University of Hamburg, Gojenbergsweg 112, 21029 Hamburg, Germany\\
$^{5}$ School of Physical Sciences, The Open University, Walton Hall, Milton Keynes, MK7 6AA, UK\\ 
$^{6}$ CSIRO Astronomy and Space Science, 26 Dick Perry Avenue, Kensington, Perth, 6151, WA, Australia\\
$^{7}$ Tor\'n Centre for Astronomy, Faculty of Physics, Astronomy and Informatics, NCU, Grudziacka 5, 87-100 Toru\'n , Poland\\
$^{8}$ Leiden Observatory, Leiden University, PO Box 9513, NL-2300 RA Leiden\\
$^{9}$ GEPI, Observatoire de Paris, CNRS, Université Paris Diderot, 5 place Jules Janssen, 92190 Meudon, France\\ 
$^{10}$ Department of Physics \& Electronics, Rhodes University, PO Box 94, Grahamstown, 6140, South Africa
}

\date{Accepted 2017 December 28. Received 2017 December 15; in original form 2017 September 5}

\pubyear{2017}

\begin{document}
\label{firstpage}
\pagerange{\pageref{firstpage}--\pageref{lastpage}}
\maketitle

\begin{abstract}
Only a small fraction of observed Active Galactic Nuclei display large-scale radio emission associated with jets, yet these radio-loud AGN have become increasingly important in models of galaxy evolution. In determining the dynamics and energetics of the radio sources over cosmic time, a key question concerns what happens when their jets switch off. The resulting `remnant' radio-loud AGN have been surprisingly evasive in past radio surveys, and therefore statistical information on the population of radio-loud AGN in their dying phase is limited. In this paper, with the recent developments of LOFAR and the VLA, we are able to provide a systematically selected sample of remnant radio-loud AGN in the Herschel-ATLAS field. Using a simple core-detection method, we constrain the upper limit on the fraction of remnants in our radio-loud AGN sample to 9 per cent, implying that the extended lobe emission fades rapidly once the core/jets turn off. We also find that our remnant sample has a wide range of spectral indices ($-1.5\leqslant \alpha^{1400}_{150}\leqslant -0.5$), confirming that the lobes of some remnants may possess flat spectra at low frequencies  just as active sources do. We suggest that, even with the unprecedented sensitivity of LOFAR, our sample may still only contain the youngest of the remnant population. 
\end{abstract}




\section{Introduction} \label{intro}
\subsection{Radio-loud AGN and their evolution}
The large number of Active Galactic Nuclei (AGN) in flux-limited radio
surveys has led to valuable statistical information on the population of radio-loud AGN \citep[e.g.][]{best07,best12}. Multi-wavelength observations of radio-loud AGN, capturing the
large-scale radio lobes inflated by the jets, have also given
important constraints on the dynamics and energetics of their
extended emission and the effects on their surrounding environment
\citep{hard02,cros04,cros05}. These observations provide tests of
(semi-)analytical and numerical models describing the time evolution
of powerful FR-II radio galaxies \citep{FR74} and their environmental impact \citep[e.g.][]{kais97,blun99,lain02,hakr13}. In particular, it has been suggested that powerful sources in
dense, cluster environments re-heat their
surrounding medium as a mechanism to offset cluster cooling -- solving
the well known `cooling flow' problem in clusters of galaxies
\citep{bass03,dunn06,mcna07,cros08}, although direct evidence of this process
for the population of radio-loud AGN in the most massive clusters
remains elusive \citep{best07}. Moreover, it has been suggested that
radio-loud AGN disrupt star formation in their host galaxies through
gas heating \citep{best06,wyle16}, regulating the growth of galaxies
through the well known AGN feedback cycle \citep[see review
  by][]{fabi12}. These advances have effectively driven the
development of a clear picture describing the evolution of radio
galaxies throughout their `active' lifetimes.

However, much is still not understood regarding the nature of
\textit{remnant AGN}, i.e the phase of the AGN life cycle that begins
when the jets switch off. Over cosmic time, AGN are expected to have
episodic nature \citep{saik09,morg17}. Clear evidence of episodic activity has come from
observations of radio galaxies with two pairs of radio lobes denoting
the inner currently active radio lobes, and the faint outer radio
lobes from previous activity, termed `double-double' radio galaxies \citep[DDRGs;][]{roet94,subr96,scho00,sari02,sari03}. Only a few such sources have been observed
\citep[e.g.][]{broc11,nand12,orru15}, and this scarcity can be explained by a
model in which the currently active radio jets drive into the
pre-existing plasma on relatively short time-scales, and therefore
merge with the outer remnant lobes \citep{kona13}. Remnant radio-loud AGN, with no evidence for currently active jets, are surprisingly even rarer than
DDRGs \citep{parm07,murg11,sari12}. During the remnant phase, the bright
radio core and jets are expected to disappear, while the lobes may
radiate for some time \citep{slee01}. Roughly half
of the energy ever transported through the jets may remain in the
radio-emitting lobes when the jets switch off (\citealt{hakr13}).
Whether or not this energy remains in the lobes in the post switch-off
phase, for how long it remains there and where the plasma ends up are
questions of great interest in studies of the AGN duty cycle.

Knowledge of the dynamics of remnants will also aid analytical models
describing the dynamical evolution of the radio plasma in the post
switch-off phase. From a physical point of view, as an active radio
galaxy expands into its environment, the lobe plasma experiences
adiabatic and radiative losses. Meanwhile, the jet continuously
replenishes the lobes with new and young electrons which will
eventually go through the same loss cycles. The fact that remnant sources have a very low detection rate \citep[e.g.][]{giov88} can be explained if the radio lobes quickly become
undetectable due to rapid energy losses\footnote{It is interesting to note that inverse-Compton emission due to the up-scattering of CMB photons by the ageing low-frequency electrons in the lobes can last for a significantly longer period than synchrotron losses at low redshift -- giving rise to IC `ghosts' seen in the X-ray \citep{mocz11}. Deep X-ray surveys may be able to find remnants previously undetected at radio-wavelengths through this method, although a sample is yet to be developed.} without the replenishment of
fresh electrons by an active jet, as well as adiabatic losses as the
lobes continue to expand. Recent analytical modelling by \cite{godf17} and \cite{brie17} have shown that models of radio galaxy evolution that consider only radiative losses over-predict the number of observed ultra-steep spectrum remnants by at least a factor of two. This further supports the idea that expansion losses in radio galaxies are important in the remnant phase, which might explain why they may quickly escape detection in current
flux-limited radio surveys.

It is therefore important to constrain the fading time of remnants from observations.  The number of remnants in the sky relative to the number of active sources in a single observation will give us some measure of this, leading to 
a robust, systematic study of remnant radio-loud AGN as a population. Such a sample will  provide a unique opportunity to understand their dynamics, and also aid the development of
analytical and numerical models describing the duty cycle of
radio-loud AGN -- an accurate description of which is currently missing in galaxy evolution models.
\subsection{Remnant selection methods}
The scarcity of continuum observations of remnant AGN has previously restricted the possibility of a detailed study in a statistical sense. Remnants are expected to be steep-spectrum in nature even
below 1 GHz (\citealt{giac07,parm07,murg11}), owing to the radiative
cooling of the lobe plasma without the input of high energy electrons
by a jet. However, solely selecting low-frequency
steep-spectrum sources (e.g. $\alpha^{1400}_{150}\leqslant -1.2$, defining the spectral index as the slope of the radio spectrum between two given frequencies in the sense $S\propto\nu^{\alpha}$) may be biased in the sense that some remnant
sources may in fact show flatter spectra \citep{brie16}, while it is
also possible for active radio galaxies to possess similarly steep
low-frequency spectra \citep{harwood13,harwood15}, which may
contaminate the remnant sample. Moreover, \cite{godf17} have shown that spectral selection methods preferentially select the oldest remnants -- a fraction of the remnant population which may form an unrepresentative sample. Morphological  criteria have
also been suggested to select remnants \citep{sari12}, since the ageing large scale
lobe emission should generally show a relaxed structure \citep{sari12},
although recently switched-off FR-II sources can contain bright
hotspots (for as long as the light travel time of the radio
galaxy -- $\ga 1$ Myr for the largest sources), and display
less relaxed and more energetic lobes that are typical of active
sources \citep[e.g. 3C28; ][]{harwood15}. \cite{brie16} suggest that morphological, spectral index and other selection methods should be used in conjunction to
give a systematic and reliable sample of the remnant population,
although multi-frequency observations at comparable
resolution can be difficult to obtain for a large sample of sources. 

The absence of a core associated with nuclear activity in radio images
is a clear signature of a remnant radio galaxy \\ 
\citep{giov88}. All genuine
remnants are expected to lack a visible radio core at all frequencies -- the fading  time for the jets, which are visible as the core on pc-scales, is as long as the light travel time of material through the jet ($\sim 10^{4}$ yr), which is a small fraction of the typical fading time of the large scale radio lobes.
Recently, \cite{godf17} have used the absence of a core in conjunction with ultra-steep spectra as a method to select a sample of candidate remnant FR-II radio galaxies with a flux density limit of $>1.5$ Jy from the 74 MHz VLA Low-Frequency Sky Survey Redux catalogue (VLSSr). Two per cent of sources were selected using their sampling criteria as candidate remnants, although this is expected to only be an upper limit, since the steep-spectrum criterion is expected to contaminate the remnant sample with high-z active sources. Selecting sources with low surface-brightness lobes in the absence of a core could form an
alternative, unbiased selection criterion.

However, it is difficult to
observe faint lobe emission in radio surveys with a high flux limit, as they are sensitive to the brightest radio sources \citep[e.g 3CRR;][]{3crr}, meaning that many remnants may be missed by the
sensitivity limits of these surveys. Detailed studies of remnants in the
past have been limited to individual sources (e.g. BLOB1,
\cite{brie16}; PKS B1400-33, \cite{subr03}; B2 0258+35, \cite{shul12}). Sensitive, low-frequency radio observations over large
areas of the sky are required to observe a larger fraction of the faint radio galaxy population associated with the remnant
phase. A robust detection of a core will  differentiate sources
between being remnant and active radio galaxies, and hence this method
is expected to provide an unbiased sample of remnants. 
\subsection{Selecting remnants with LOFAR}
Faint radio lobes from remnant radio-loud AGN are expected to be
detected with the Low-frequency Array (LOFAR; \citealt{vanh13}) operating at around 150 MHz, where the ageing radio lobes would be
brighter than at GHz frequencies due to the preferential cooling of
higher-energy electrons \citep{kard62,pach70}. Crucially, LOFAR has the
resolution ($\sim$6 arcsec at 150 MHz with the full Dutch array) and
$uv$-coverage to make sensitive and deep surveys of the sky. The
mixture of long and short baselines in a single observation and the wide field of view enables a
large number of potential remnant sources to be detected. LOFAR thus
gives us the opportunity to produce a much-needed systematic survey of
the remnant radio-loud AGN population.

\cite{brie17} have recently made use of LOFAR observations of the Lockman Hole field to assess the efficiency of various spectral and morphological criteria in selecting remnant radio galaxies. An initial sample of extended radio galaxies from the LOFAR images was obtained, and cross-matching the sources with the Faint Images of the Radio Sky at Twenty-cm \citep[FIRST;][]{beck95} survey was performed. Those sources that did not show significant evidence of a radio core based on the FIRST images were deemed  candidate remnants. Interestingly, the remnant fraction obtained in this sample was $30$ per cent. This remnant fraction, if robust, would be considerably higher than that found in the 3CRR sample, which would strengthen the original hypothesis that remnants are more
detectable with sensitive telescopes such as LOFAR. Moreover, it implies a much longer remnant lifetime than that
generally assumed, potentially giving important constraints on the
energy loss processes in these sources. Further samples of candidate remnants obtained using this method would make this result more robust.

An exploratory LOFAR High-Band
Antenna (HBA, 110-200 MHz) survey of the \textit{Herschel}-ATLAS North
Galactic Pole (H-ATLAS NGP) field has also been carried out recently \citep[][hereafter H16]{hard_lofar16}. This sky region covers
approximately 142 square degrees centred around RA $= 13.5$ h and
DEC$=30$\textdegree . Around 15,000 discrete radio sources were
detected using observations between 126-173 MHz using the full
Dutch array, giving a resolution of $\sim 10 \times6$ arcsec. For
identification of the radio sources in this sample, visual cross-matching
with the Sloan Digital Sky Survey Data Release 12 \citep[SDSS DR12;][]{alam15} at $r$-band to find the most likely
optical host galaxy, assisted by any 1.4 GHz  FIRST core detection, was performed. For sources without FIRST core detections, IDs were made based on the morphology of the LOFAR emission alone. Bright radio sources associated with star formation, which
are indistinguishable from radio-loud AGN based on the LOFAR data
alone, were separated from the sample using the FIR-radio correlation
aided by \textit{Herschel} data. See \citetalias{hard_lofar16} for full details of
  observations, data calibration and AGN/star formation separation
  techniques.

As an initial search for potential remnants, a parent sample of radio-loud AGN selected from the LOFAR observations was made satisfying the following
criteria: bright (>80 mJy at 150 MHz),  ensuring that the sample is
flux-complete while also allowing a faint but active core to be
detected from follow-up higher-resolution observations; well resolved
(>40 arcsec) to show extended emission; and classed as AGN based on
the FIR-radio correlation. Of $127$ such sources, $38$ sources were
then selected as \textit{candidate} remnants on the basis that there
was no evidence of a core detected with FIRST -- giving a potential
remnant fraction of $30$ per cent, in agreement with \cite{brie17}. 

However, the remnant fraction obtained from LOFAR is limited by the
fact that FIRST is not particularly sensitive to the detection of
compact cores, meaning that the remnant fractions of
\citetalias{hard_lofar16} and \cite{brie17} must be regarded as only an upper limit. At a
resolution of $\sim 5$ arcsec, FIRST has a typical source detection threshold of
1 mJy. However, if
we define the core prominence as the ratio of the FIRST core flux density at
1.4 GHz and the total flux density at 150 MHz (the exact frequency for
the measurement of a core flux density is unimportant, since the cores
of radio-loud AGN are expected to be flat-spectrum out to high
frequencies \citep{halo08}), then the 3$\sigma$ upper limit for the
core prominence for the faintest objects in the sample detected with
FIRST is $0.4/80\approx 5\times 10^{-3}$. On the other hand, the
\textit{median} value of core prominence in the brighter 3CRR sample
is $\sim 3\times 10^{-4}$ \citep{mull08}. Although 3CRR selects for
the brightest sources, and so would be expected to have systematically
low core prominences relative to that of the LOFAR sample, it is still
clearly possible that faint radio cores will be missed in the LOFAR
sample if only FIRST is used to identify them. To decide whether this is simply a result of the sensitivity limits of
FIRST, or of these sources actually having no radio-bright cores
denoting nuclear activity, requires more sensitive and
higher-resolution observations. Furthermore, without the detection of
a radio core, optical ID cross-matching clearly becomes a
challenge, and therefore we cannot rely on the optical ID for sources
without a FIRST-detected core.

In this paper, we present new 6 GHz Karl G. Jansky Very Large Array
(VLA) snapshot observations of the 38 LOFAR candidate remnants
obtained by \citetalias{hard_lofar16} (sources listed in Table \ref{targetinfo}). With a higher spatial resolution ($\sim$ 0.3
arcsec for the longest baselines at 6 GHz) and an order of magnitude
improvement in sensitivity than that of FIRST, we can obtain more
robust detections/non-detections of a core. A clear detection of a
core in the new 6 GHz images will mean that the corresponding source
is not a remnant, and conversely a non-detection of a core will retain
the candidate remnant status for the source, thereby giving a more
accurate constraint on the remnant fraction. We will then be able to
provide a systematic sample of the candidate remnant population in the
NGP H-ATLAS field, leading to important statistical information
regarding the population of remnant radio-loud AGN. Furthermore, the detection of a compact radio core with the VLA at
the position of the host galaxy of the source will enable previously
ambiguous optical identifications to be confirmed or rejected. 

The remainder of this paper is structured as follows. In Section \ref{jvladata}, we
describe the VLA observations and data reduction processes for the 38
remnant candidate sources from the H-ATLAS NGP field. In Section \ref{radioimages} we
present the images, followed by a qualitative analysis of each candidate source,
outlining their remnant or active status, as well as any optical
misidentification, throughout Section \ref{results}. We discuss the newly constrained remnant fraction
in Section \ref{discussion}, and its implications for the dynamical evolution of
radio-loud AGN. We then conclude with a brief summary of the results
in Section \ref{conclusions}.

Throughout this work we use cosmological parameters
based on a $\Lambda$CDM cosmology with $H_0=70$ km s$^{-1}$Mpc$^{-1}$,
$\Omega_m=0.3$ and $\Omega_{\Lambda}=0.7$. Co-ordinate positions are
given in the J2000 system. Spectral indices are defined in the sense $S_{\nu}\propto \nu^{\alpha}$.
\section{Observations}\label{observations}
\subsection{VLA Data Reduction}\label{jvladata}
The 38 remnant candidates were observed  with the VLA on the 30th
September 2016 in the A-configuration and on the 8th September 2016 in
the B-configuration (Table \ref{observations}). Observations were made in the broad-band C-band
system in 3-bit mode (4 GHz bandwidth from 4-8 GHz). Since we only
required snapshot observations, exposure times were $\sim 5$ minutes
per source, reaching background rms levels of $\sim 10-15 \mu$Jy/beam in
the final combined images -- an improvement in sensitivity to a
radio core by an order of magnitude over FIRST. The observations were targeted at the SDSS optical ID for each source (given by
\citetalias{hard_lofar16}). The sources lie in a roughly $12 \times
12$-degree region of the sky, including the quasar 3C286 which was
used as the flux calibrator. The blazar J1310+3233 was used as the
complex gain calibrator, observed 13 times between different scans of
target sources. These details are summarised in Table \ref{jvlaobs}.

Prior to calibration the {\sc
  AOFlagger} software (\citealt{offr12}) was
used on the data to automatically flag radio-frequency interference
(RFI). The data were then reduced using the {\sc CASA} VLA pipeline version
1.3.5 for reduction using {\sc CASA} version 4.5.0 \citep{mcmu07}. A
selection of gain calibration tables were inspected to check the
quality of the calibration and/or the occurrence of bad baselines that
were not automatically flagged previously, prior to imaging. As a
secondary flagging process, the {\sc CASA} tool `rflag' was also applied to
both measurement sets to remove additional RFI present in the $uv$
data. Initial images were then produced by  CLEANing \citep{hogbom}
both A and B-configuration measurement sets separately for each target
source, using the imaging parameters detailed in Table \ref{casaparams}. For J131040.25+322044.1 (Figure \subref{J131040.03+322}), a bright quasar in the field prevented suitable deconvolution and resulted in high RMS noise levels in the vicinity of the optical ID. For this source only, the data were calibrated manually, and imaged using the imager {\sc WSClean} \citep{offr14} to reduce errors due to poor deconvolution in the wide-field image.
Self-calibration was possible for some of the targets with strong core
detections, and in the event of non-detections, bright sources
present nearby in the field were used.  The individually self-calibrated A and B-configuration data sets were
then combined in the $uv$ plane, producing the final radio maps with
an average background noise level of $\sim 14.6 \ \mu$Jy/beam, shown in
Figure \ref{jvlaimages}. For presentation purposes, we  smoothed the VLA maps with a
Gaussian function having a FWHM of 0.2 arcsec ($\sim 5$ pixels across).

A software problem with the VLA regarding the delay model used for
correlation between antennas potentially introduced a problem into
observations made between 9th August and 14th November 2016 in
the A-configuration. The main effect of this subtle bug would have
been a displacement of a source's position in the direction of
elevation. This might have affected our analysis of the data since our
science aims depend on our ability to confirm or reject optical IDs
based on the position of a VLA core detection, although a potential
source offset was only deemed to be serious for sources below a
declination of 20 degrees and all of our sources are in fact at higher
declination (Table \ref{targetinfo}). Nevertheless we estimated the magnitude of the
potential offset for each target source, and for each source where the
calculated offset was larger than the beam-size (5/38 sources), we
visually inspected the A-configuration image overlaid on the
corresponding SDSS image. We found that, for those targets with
detected cores, there was a positional match with the SDSS optical ID.
For those without cores, we ensured that the magnitude of the
potential offset would not have allowed the target source to be
associated with a different optical host. We concluded
that this software problem was not manifested in our observations at a
level that would have required re-observation. 

\subsection{LOFAR data}\label{lofardata}
The LOFAR data used are those presented by \citetalias{hard_lofar16}, but
for this paper we use a new direction-dependent calibration procedure.
This processing of the H-ATLAS data will be described in more detail
elsewhere but, to summarize briefly, it involves replacing the facet
calibration method described by \citetalias{hard_lofar16} with a
direction-dependent calibration using the methods of
\cite{Tasse14b,Tasse14a}, implemented in the software package {\sc
  killms}, followed by imaging with a newly developed imager {\sc
  ddfacet} (Tasse et al.\ 2017 in prep.) which is capable of applying
these direction-dependent calibrations in the process of imaging. The
H-ATLAS data were processed using the December 2016 version of the
pipeline, {\sc ddf-pipeline}\footnote{See
  \url{http://github.com/mhardcastle/ddf-pipeline} for the code.},
that is under development for the processing of the LoTSS survey
\citep[][and in prep.]{Shimwell+17}. The main advantage of this
reprocessing is that it gives lower noise and higher image fidelity
than the process described by \citetalias{hard_lofar16}, increasing the
point-source sensitivity and removing artefacts from the data, but it
also allows us to image at a slightly higher resolution -- the images
used in this paper have a 7-arcsec restoring beam. Note that, due to the increased point-source sensitivity and reduction in noise, the most faint and diffuse sources of emission are less well represented than in the previous images. Therefore, for sources J125422.44+304 (Figure \subref{J125422.44+304}),
J132602.42+314 (Figure \subref{J132602.42+314}) and J132622.56+320
(Figure \subref{J132622.56+320}), we use the pre-processed LOFAR data presented by \citetalias{hard_lofar16}, for a better visual  representation of the source structure. We ensured that these sources still met our sample criteria detailed in Section \ref{intro}, based on the new re-processed LOFAR data.

After careful visual inspection
of the re-processed LOFAR images, we concluded that sources 
J130849.75+252 (Figure \subref{J130849.75+252}), J131537.33+310 (Figure
\subref{J131537.33+310}), J131828.97+291 (Figure
\subref{J131828.97+291}) and J131832.33+291
(Figure \subref{J131832.33+291}) are either  imaging artefacts that were
misidentified as real sources based on the lower resolution LOFAR
images presented by \citetalias{hard_lofar16}, or real sources that now do not meet our flux and angular extent sample criteria. We therefore  removed these sources from our sample, leaving our parent AGN sample at 123 sources, and our candidate remnant sample at 34 sources.
\begin{table*}
\caption{Candidate remnant sources. SDSS ID gives the name of the SDSS source corresponding to the optical host. RA and DEC specify the co-ordinates of this optical ID (also the location of VLA pointing for that source). LOFAR flux density gives the flux density of the radio source at 150 MHz associated with the optical ID. Redshifts are either photometric (p) or spectroscopic (s), as stated below, and where both were available the spectroscopic redshift is given.}
\centering
\begin{tabular}{cccccccc}
\hline
\vspace{1mm}
LOFAR name & SDSS ID & \multicolumn{1}{p{1.5cm}}{\centering RA \\ (h:m:s)} & \multicolumn{1}{p{1.5cm}}{\centering DEC \\ ($^{\circ}$:$'$:$''$)} & \multicolumn{1}{p{1.5cm}}{\centering LOFAR 150 MHz flux density (Jy)} & \multicolumn{1}{p{1.5cm}}{\centering Redshift \\ ($z$)} & \multicolumn{1}{p{1.5cm}}{\centering Redshift type \\ (p/s)}\\
\hline
J125143.00+332020.2 & 1237665228382077637 &  12:51:42.90 & +33:20:20.66 & 0.17843 & 0.3866 & p\\
J125147.02+314046.0 & 1237665226234659106 &  12:51:47.02 & +31:40:47.66 & 0.21616 & 0.3579 & s\\
J125311.08+304029.2 & 1237665443125723293 &  12:53:11.62 & +30:40:17.35 & 0.09572 & 0.3497 & s\\
J125419.40+304803.0 & 1237667255629643949 &  12:54:22.44 & +30:47:28.10 & 0.28814 & 0.1949 & p\\
J125931.85+333654.2 & 1237665023834915425 &  12:59:30.80 & +33:36:46.96 & 0.11063 & 0.5367 & p\\
J130003.72+263652.1 & 1237667442438111406 &  13:00:04.24 & +26:36:52.70 & 0.18476 & 0.3164 & s\\
J130013.55+273548.7 & 1237667323797766776 &  13:00:14.65 & +27:36:00.07 & 0.12813 & 0.3228 & p\\
J130332.38+312947.1 & 1237665226235773388 &  13:03:32.47 & +31:29:49.54 & 0.08529 & 0.7412 & s\\
J130415.46+225322.6 & 1237667736123933309 &  13:04:15.96 & +22:53:43.49 & 0.26408 & 0.3723 & p\\
J130532.49+315639.0 & 1237665226772775688 &  13:05:32.01 & +31:56:34.86 & 0.13407 & 0.3318 & p\\
J130548.65+344052.7 & 1237665025446052014 &  13:05:48.81 & +34:40:53.84 & 0.19518 & 0.2318 & p\\
J130640.99+233824.8 & 1237667446197649528 &  13:06:41.12 & +23:38:23.49 & 0.16415 & 0.1829 & s\\
J130825.47+330508.2 & 1237665228383650573 &  13:08:26.27 & +33:05:15.05 & 0.09703 & 0.4298 & p\\
J130849.22+252841.8 & 1237667912745550347 &  13:08:49.74 & +25:28:40.23 & 0.08652 & 0.4157 & p\\
J130916.85+305118.3 & 1237665225699492186 &  13:09:16.02 & +30:51:21.94 & 0.11871 & 0.3459 & p\\
J130915.61+230309.5 & 1237667783913439350 &  13:09:16.66 & +23:03:11.37 & 0.18042 & 0.2281 & s\\
J130917.63+333028.4 & 1237665023835833039 &  13:09:17.73 & +33:30:35.68 & 0.27835 & 0.4922 & p\\
J131040.25+322044.1 & 1237665330930778227 &  13:10:40.03 & +32:20:47.66 & 0.33559 & 0.5517 & s\\
J131039.92+265111.9 & 1237667442976030877 &  13:10:40.51 & +26:51:05.49 & 0.08053 & 0.1853 & s\\
J131235.35+331348.6 & 1237665126938117173 &  13:12:36.14 & +33:13:38.99 & 0.08448 & 0.7846 & p\\
J131405.16+243234.1 & 1237667911672333046 &  13:14:05.90 & +24:32:40.38 & 0.25808 & 0.3806 & p\\
J131446.57+252819.8 & 1237667448882725588 &  13:14:46.82 & +25:28:20.51 & 0.10939 & 0.5438 & s\\
J131536.30+310615.6 & 1237665226236887706 &  13:15:37.33 & +31:06:15.61 & 0.09244 & 0.2775 & p\\
J131827.83+291658.5 & 1237665442054471904 &  13:18:28.96 & +29:17:26.48 & 0.13550 & 0.3027 & p\\
J131833.81+291904.9 & 1237665442054471911 &  13:18:32.32 & +29:18:38.94 & 0.08778 & 0.2705 & p\\
J132402.51+302830.1 & 1237665225700868124 &  13:24:03.44 & +30:28:22.86 & 0.35482 & 0.0488 & p\\
J132602.06+314645.6 & 1237665227311546592 &  13:26:02.42 & +31:46:50.42 & 0.12590 & 0.2370 & s\\
J132622.12+320512.1 & 1237665227848482831 &  13:26:22.56 & +32:05:02.36 & 0.16374 & 0.3489 & p\\
J132738.77+350644.6 & 1237664671644517294 &  13:27:35.32 & +35:06:36.73 & 0.83581 & 0.5003 & p\\
J132949.25+335136.2 & 1237665128550236906 &  13:29:48.87 & +33:51:52.22 & 0.14627 & 0.5601 & s\\
J133016.12+315923.9 & 1237665227848811145 &  13:30:16.32 & +31:59:19.77 & 0.20335 & 0.3106 & p\\
J133058.91+351658.9 & 1237664852028948635 &  13:30:57.33 & +35:16:50.29 & 0.22973 & 0.3158 & s\\
J133309.94+251045.2 & 1237667321653625933 &  13:33:10.56 & +25:10:44.12 & 0.10982 & 0.2639 & p\\
J133422.15+343640.5 & 1237665129624371524 &  13:34:22.21 & +34:36:34.79 & 0.13971 & 0.5575 & p\\
J133502.36+323312.8 & 1237665023838257509 &  13:35:02.38 & +32:33:13.94 & 0.24344 & 0.4282 & p\\
J133642.53+352009.9 & 1237664852566278355 &  13:36:43.09 & +35:20:11.72 & 0.13395 & 0.1145 & s\\
J134702.03+310913.3 & 1237665330934186147 &  13:47:01.71 & +31:09:24.21 & 0.28446 & 0.1995 & p\\
J134802.83+322938.3 & 1237665024376307813 &  13:48:02.70 & +32:29:40.10 & 0.13825 & 0.2101 & s\\
\hline
\end{tabular}
\label{targetinfo}
\end{table*}
\begin{table*}
\caption{VLA observation details}
\centering
\begin{threeparttable}
\begin{tabular}{ccccccc}
\hline
VLA project code & Array & Bandwidth (GHz) & Obs. date & Duration & Flux calibrator & Average RMS ($\mu$Jy/beam) \\
\hline
16B-245 & A & 4-8 & 30/09/16 & 3.4h & 3C286 & 19.9\tnote{1}\\
16B-245 & B & 4-8 & 08/09/16 & 3.4h & 3C286 & 19.9\tnote{1}\\
\hline
\end{tabular}
\begin{tablenotes}
\item[1] Background RMS taken from radio maps prior to self-calibration, and therefore some maps with bright background sources contain high noise levels.
\end{tablenotes}
\end{threeparttable}
\label{jvlaobs}
\end{table*}
\begin{table*}
\caption{Summary of CASA imaging parameters for the 6 GHz VLA observations. Shown are the CLEAN parameters used to image the visibilities from both A- and B-configuration measurement sets.}
\centering
\begin{threeparttable}
\begin{tabular}{llll}
\hline
Parameter & CASA name & Value & Units\\
\hline \vspace{0.1mm}
Cell size & cellsize & 0.04$\times$0.04 & arcsec\\
Image size\tnote{1} & imsize & 4096$\times$4096 & pixels\\
Noise threshold & noise & 0.01 & mJy/beam\\
Weighting\tnote{2} & robust & 0.0 & \\
Average beam major axis\tnote{3} & BMAJ & 0.47 & arcsec\\
Average beam minor axis\tnote{3} & BMIN & 0.28 & arcsec \\ 
\hline
\end{tabular}
\begin{tablenotes}
\item[1] The images shown in Figures \subref{J125422.44+304} and  \subref{J132735.32+350} have a image size of $8192\times 8192$, due to the large angular extent of the sources.
\item[2] Standard `Briggs' weighting characterised by `robust' parameter.
\item[3] Beam size taken as an average from  the 38 scans of the target sources.
\end{tablenotes}
\end{threeparttable}
\label{casaparams}
\end{table*}
\subsection{Radio images}\label{radioimages}
\begin{figure*}
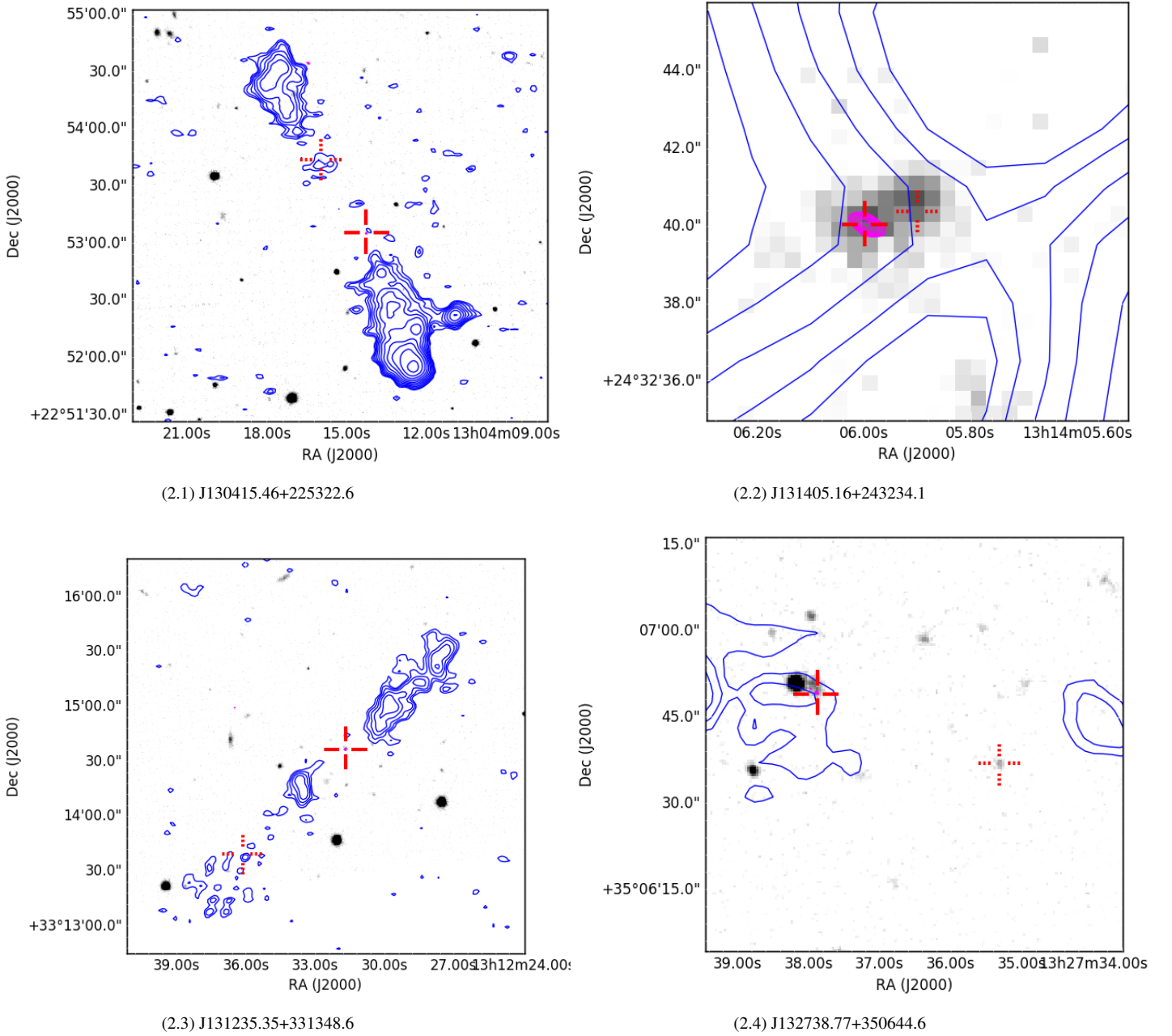

\centering
\begin{subfigure}[b]{0.475\textwidth}
 	\centering        		  \includegraphics[width=9.25cm]{{{new_opticalID/Field13_J130415.97+225}}}\\
    \caption{J130415.46+225322.6}
    \label{newopticalidJ130415.97+225}
\end{subfigure}
\hfill
\begin{subfigure}[b]{0.475\textwidth}
	\centering
    \includegraphics[width=9.25cm]{{{new_opticalID/Field17_J131405.90+243}}}\\
    \caption{J131405.16+243234.1}
    \label{newopticalidJ131405.90+243}
\end{subfigure}
\vskip\baselineskip
\begin{subfigure}[b]{0.475\textwidth}
	\centering
    \includegraphics[width=9.5cm]{{{new_opticalID/Field27_J131236.14+331}}}\\
    \caption{J131235.35+331348.6}
    \label{newopticalidJ131236.14+331}
\end{subfigure}
\hfill
\begin{subfigure}[b]{0.475\textwidth}
	\centering
    \includegraphics[width=9.5cm]{{{new_opticalID/Field39_J132735.32+350}}}\\
    \caption{J132738.77+350644.6}
    \label{newopticalidJ132735.32+350}
\end{subfigure} 
\caption{SDSS $r$-band images, shown in grey-scale, of the four new optical identifications made in this sample. Overlaid are LOFAR 150-MHz contours (blue) and VLA 6-GHz contours (magenta). For clarity, the previous optical ID is marked  with a red dotted cross-hair, and the new optical ID we make based on the position of the VLA 6 GHz core is marked with a red solid cross-hair.}
\label{newopticalid}
\end{figure*}

In Figure \ref{jvlaimages} we present the VLA 6 GHz images overlaid with the LOFAR 150-MHz contours (contour levels at increasing powers of 2$\sigma$) in order to view the faint low-frequency plasma of the candidate remnants, or the lobes of potentially active sources. To emphasise the visibility of a potential faint and/or compact radio core, we provide a zoom-in of the VLA image at the location of the current optical ID (marked with a cross-hair) on the top-right hand side of each image. We define all 3$\sigma$ core detections at the optical ID to be active sources. Note that the non-detection of a core ($<3\sigma$) at the optical ID as seen by these zoom-ins mean that the source may have been optically misidentified, and that there may be a faint active core elsewhere in the 6 GHz map at a location corresponding to a different optical ID associated with the radio source. In this case, we confirm the active nature of the source and also propose a new SDSS ID, as displayed in Figure \ref{newopticalid} and detailed in Table \ref{opticalmisidtable}. Otherwise, in the absence of a core, we keep the candidate remnant status for the source as given by \citetalias{hard_lofar16}. Physical descriptions of each source, along with confirmation of remnant statuses and optical IDs, based on these images, are given in Section \ref{remnantstatus}. The four sources that we removed from the candidate remnant sample of \citetalias{hard_lofar16} are also shown in Figure \ref{jvlaimages}, and are labelled in their figure caption. Note that in the case of confirming a new optical ID, and hence a different redshift for the source, the physical scale bar on the bottom right of each image is incorrect. We correct these physical parameters for these few sources in our analysis (Section \ref{discussion}).
\begin{figure*}
\centering
\begin{subfigure}[b]{0.475\textwidth}
	\centering    
    \includegraphics[width=9.5cm]{{{Field8_J125142.91+332}}}\\
    \caption{J125143.00+332020.2}
    \label{J125142.91+332}
\end{subfigure}
\hfill
\begin{subfigure}[b]{0.475\textwidth}
	\centering    
    \includegraphics[width=9.5cm]{{{Field5_J125147.03+314}}}\\
    \caption{J125147.02+314046.0}
    \label{J125147.03+314}
\end{subfigure} 
\vskip\baselineskip
\begin{subfigure}[b]{0.475\textwidth}
	\centering    
    \includegraphics[width=9.5cm]{{{Field6_J125311.62+304}}}\\
    \caption{J125311.08+304029.2}
    \label{J125311.62+304}
\end{subfigure}
\hfill
\begin{subfigure}[b]{0.475\textwidth}
	\centering    
    \includegraphics[width=9.5cm]{{{Field7_J125422.44+304}}}\\
    \caption{J125419.40+304803.0$^{\dagger}$}
    \label{J125422.44+304}
\end{subfigure}
\caption{6-GHz VLA images of the candidate remnant sources (shown as the background), centred on the optical IDs made by \protect\citetalias{hard_lofar16}. A zoom-in of the VLA image at the location of the optical ID is given on the top right hand side of each image, for a clear visual identification of a core. Images are scaled logarithmically, and smoothed with a Gaussian function with FWHM of 0.2 arcsec (5 pixels). 150-MHz LOFAR contours (blue) are set at various powers of 2 multiplied by 2$\sigma$ (where $\sigma$ is defined as the local noise level in the LOFAR image), in order to encapsulate as much low surface brightness emission as possible. Sources that we identify as artefacts and not radio-loud AGN are labelled in their figure caption. $^{\dagger}$Contours based on the pre-processed LOFAR data (see Section \ref{lofardata}).}
\label{jvlaimages}
\end{figure*}
\begin{figure*}\ContinuedFloat
\centering
\begin{subfigure}[b]{0.475\textwidth}
	\centering    
    \includegraphics[width=9.5cm]{{{Field9_J125930.81+333}}}\\
    \caption{J125931.85+333654.2}
    \label{J125930.81+333}
\end{subfigure}
\hfill
\begin{subfigure}[b]{0.475\textwidth}
	\centering    
    \includegraphics[width=9.5cm]{{{Field12_J130004.25+263}}}\\
    \caption{J130003.72+263652.1} 
	\label{J130004.25+263}
\end{subfigure}
\vskip\baselineskip
\begin{subfigure}[b]{0.475\textwidth}
	\centering    
    \includegraphics[width=9.5cm]{{{Field11_J130014.65+273}}}\\
    \caption{J130013.55+273548.7}
    \label{J130014.65+273}
\end{subfigure}
\hfill
\begin{subfigure}[b]{0.475\textwidth}
	\centering    
    \includegraphics[width=9.5cm]{{{Field22_J130332.47+312}}}\\
    \caption{J130332.38+312947.1}
    \label{J130332.47+312}
\end{subfigure}
\vskip\baselineskip
\begin{subfigure}[b]{0.475\textwidth}
	\centering    
    \includegraphics[width=9.5cm]{{{Field13_J130415.97+225}}}\\
    \caption{J130415.46+225322.6}
    \label{J130415.97+225}
\end{subfigure}
\hfill
\begin{subfigure}[b]{0.475\textwidth}
	\centering   
    \includegraphics[width=9.5cm]{{{Field23_J130532.02+315}}}\\
    \caption{J130532.49+315639.0}
    \label{J130532.02+315} 
\end{subfigure}
\caption{Continued}
\end{figure*}
\begin{figure*}\ContinuedFloat
\centering
\begin{subfigure}[b]{0.475\textwidth}
	\centering    
    \includegraphics[width=9.5cm]{{{Field10_J130548.81+344}}}\\
    \caption{J130548.65+344052.7}
    \label{J130548.81+344}
\end{subfigure}
\hfill
\begin{subfigure}[b]{0.475\textwidth}
	\centering    
    \includegraphics[width=9.5cm]{{{Field14_J130641.12+233}}}\\
    \caption{J130640.99+233824.8}
    \label{J130641.12+233}
\end{subfigure}
\vskip\baselineskip
\begin{subfigure}[b]{0.475\textwidth}
	\centering    
    \includegraphics[width=9.5cm]{{{Field25_J130826.28+330}}}\\
    \caption{J130825.47+330508.2}
    \label{J130826.28+330}
\end{subfigure}
\hfill
\begin{subfigure}[b]{0.475\textwidth}
	\centering    
    \includegraphics[width=9.5cm]{{{Field16_J130849.75+252}}}\\
    \caption{J130849.22+252841.8 (removed from sample)}
    \label{J130849.75+252}
\end{subfigure}
\vskip\baselineskip
\begin{subfigure}[b]{0.475\textwidth}
	\centering    
    \includegraphics[width=9.5cm]{{{Field24_J130916.02+305}}}\\
    \caption{J130916.85+305118.3}
    \label{J130916.02+305}
\end{subfigure}
\hfill
\begin{subfigure}[b]{0.475\textwidth}
	\centering    
    \includegraphics[width=9.5cm]{{{Field15_J130916.66+230}}}\\
    \caption{J130915.61+230309.5}
    \label{J130916.66+230}
\end{subfigure}
\caption{Continued}
\end{figure*}
\begin{figure*}\ContinuedFloat
\centering
\begin{subfigure}[b]{0.475\textwidth}
 	\centering        		  \includegraphics[width=9.5cm]{{{Field26_J130917.74+333}}}\\
    \caption{J130917.63+333028.4 (removed from sample)}
    \label{J130917.74+333}
\end{subfigure}
\hfill
\begin{subfigure}[b]{0.475\textwidth}
	\centering
    \includegraphics[width=9.5cm]{{{Field28_J131040.03+322}}}\\
    \caption{J131040.25+322044.1}
    \label{J131040.03+322}
\end{subfigure}
\vskip\baselineskip
\begin{subfigure}[b]{0.475\textwidth}
	\centering
    \includegraphics[width=9.5cm]{{{Field19_J131040.52+265}}}\\
    \caption{J131039.92+265111.9}
    \label{J131040.52+265}
\end{subfigure}
\hfill
\begin{subfigure}[b]{0.475\textwidth}
	\centering
    \includegraphics[width=9.5cm]{{{Field27_J131236.14+331}}}\\
    \caption{J131235.35+331348.6}
    \label{J131236.14+331}
\end{subfigure}
\vskip\baselineskip
\begin{subfigure}[b]{0.475\textwidth}
	\centering
    \includegraphics[width=9.5cm]{{{Field17_J131405.90+243}}}\\
    \caption{J131405.16+243234.1}
    \label{J131405.90+243}
\end{subfigure}
\hfill
\begin{subfigure}[b]{0.475\textwidth}
	\centering
    \includegraphics[width=9.5cm]{{{Field18_J131446.83+252}}}\\
    \caption{J131446.57+252819.8}
    \label{J131446.83+252}
\end{subfigure}
\caption{Continued}
\end{figure*}
\begin{figure*}\ContinuedFloat
\centering
\begin{subfigure}[b]{0.475\textwidth}
 	\centering        		  \includegraphics[width=9.5cm]{{{Field29_J131537.33+310}}}\\
    \caption{J131536.30+310615.5 (removed from sample)}
    \label{J131537.33+310}    
\end{subfigure}
\hfill
\begin{subfigure}[b]{0.475\textwidth}
	\centering
    \includegraphics[width=9.5cm]{{{Field21_J131828.97+291}}}\\
    \caption{J131827.83+291658.5 (removed from sample)}
    \label{J131828.97+291}
\end{subfigure}
\vskip\baselineskip
\begin{subfigure}[b]{0.475\textwidth}
	\centering
    \includegraphics[width=9.5cm]{{{Field20_J131832.33+291}}}\\
    \caption{J131833.81+291904.9 (removed from sample)}
    \label{J131832.33+291}
\end{subfigure}
\hfill
\begin{subfigure}[b]{0.475\textwidth}
	\centering
    \includegraphics[width=9.5cm]{{{Field30_J132403.45+302}}}\\
    \caption{J132402.51+302830.1}
    \label{J132403.45+302}
\end{subfigure}
\vskip\baselineskip
\begin{subfigure}[b]{0.475\textwidth}
	\centering
    \includegraphics[width=9.5cm]{{{Field31_J132602.42+314}}}\\
    \caption{J132602.06+314645.6$^{\dagger}$}
    \label{J132602.42+314}
\end{subfigure}
\hfill
\begin{subfigure}[b]{0.475\textwidth}
	\centering
    \includegraphics[width=9.5cm]{{{Field32_J132622.56+320}}}\\
    \caption{J132622.12+320512.1$^{\dagger}$}
    \label{J132622.56+320}
\end{subfigure}
\caption{Continued}
\end{figure*}
\begin{figure*}\ContinuedFloat
\centering
\begin{subfigure}[b]{0.475\textwidth}
 	\centering        		  \includegraphics[width=9.5cm]{{{Field39_J132735.32+350}}}\\
    \caption{J132738.77+350644.6}
    \label{J132735.32+350}
\end{subfigure}
\hfill
\begin{subfigure}[b]{0.475\textwidth}
	\centering
    \includegraphics[width=9.5cm]{{{Field35_J132948.87+335}}}\\
    \caption{J132949.25+335136.2}
    \label{J132948.87+335}
\end{subfigure}
\vskip\baselineskip
\begin{subfigure}[b]{0.475\textwidth}
	\centering
    \includegraphics[width=9.5cm]{{{Field33_J133016.33+315}}}\\
    \caption{J133016.12+315923.9}
    \label{J133016.33+315}
\end{subfigure}
\hfill
\begin{subfigure}[b]{0.475\textwidth}
	\centering
    \includegraphics[width=9.5cm]{{{Field38_J133057.34+351}}}\\
    \caption{J133058.91+351658.9}
    \label{J133057.34+351}
\end{subfigure}
\vskip\baselineskip
\begin{subfigure}[b]{0.475\textwidth}
	\centering
    \includegraphics[width=9.5cm]{{{Field1_J133310.56+251}}}\\
    \caption{J133309.94+251045.2}
    \label{J133310.56+251}
\end{subfigure}
\hfill
\begin{subfigure}[b]{0.475\textwidth}
	\centering
    \includegraphics[width=9.5cm]{{{Field36_J133422.21+343}}}\\
    \caption{J133422.15+343640.5}
    \label{J133422.21+343}
\end{subfigure}
\caption{Continued}
\end{figure*}
\begin{figure*}\ContinuedFloat
\centering
\begin{subfigure}[b]{0.475\textwidth}
 	\centering        		  \includegraphics[width=9.5cm]{{{Field34_J133502.38+323}}}\\
    \caption{J133502.36+323312.8}
    \label{J133502.38+323}
\end{subfigure}
\hfill
\begin{subfigure}[b]{0.475\textwidth}
	\centering
    \includegraphics[width=9.5cm]{{{Field37_J133643.09+352}}}\\
    \caption{J133642.53+352009.9}
    \label{J133643.09+352}
\end{subfigure}
\vskip\baselineskip
\begin{subfigure}[b]{0.475\textwidth}
	\centering
    \includegraphics[width=9.5cm]{{{Field2_J134701.71+310}}}\\
    \caption{J134702.03+310913.3}
    \label{J134701.71+310}
\end{subfigure}
\hfill
\begin{subfigure}[b]{0.475\textwidth}
	\centering
    \includegraphics[width=9.5cm]{{{Field3_J134802.70+322}}}\\
    \caption{J134802.83+322938.3}
    \label{J134802.70+322}
\end{subfigure}
\caption{Continued}
\end{figure*}
\clearpage
\section{Results}\label{results}
\subsection{Remnant statuses}\label{remnantstatus}
In this section we report on the status of each radio source -- either active or remnant. As explained in Section \ref{intro}, we use the simple definition of a remnant AGN as a radio source, classified as an AGN based on the methods of \citetalias{hard_lofar16}, without a compact radio core and jets, aided by our new 6 GHz VLA observations. Confirmations of optical IDs (previously made by \citetalias{hard_lofar16}) were carried out visually by cross matching our detected cores with catalogues from SDSS Data Release 12 \citep{alam15}, in addition to FIRST and NVSS at 1.4 GHz. Where any compact radio core from our new 6 GHz observations coincides with a different optical ID (see Figure \ref{newopticalid}), we assign the new SDSS optical ID to the radio source and give its IAU name, detailed in the descriptions of each source below and in Table \ref{opticalmisidtable}.
\subsubsection{J125143.00+332020.2: Active}
Figure \subref{J125142.91+332} shows a clear central core with what seems to be faint jet emission at 6 GHz extending to the north-east, with more extended lower frequency emission at 150 MHz on both sides of the core. A hotspot is also seen near the edge of the western lobe at 6 GHz.
\subsubsection{J125147.02+314046.0: Active}
This is a clearly active DDRG, as evident in Figure \subref{J125147.03+314}. The 150 MHz contours show the outer radio lobes indicative of previous AGN activity, while the 6 GHz emission shows the hotspots from the restarted jets as well as a central core.
\subsubsection{J125311.08+304029.2: Active}
Figure \subref{J125311.62+304} shows a clear central core, which coincides with the position of the current optical ID from the SDSS catalogue. Extended structure is also seen with at 150 MHz, although relatively faint, and extending in different axes relative to each other. It is possible that this source is in a group environment, based on the large number of SDSS sources surrounding the radio source, and that the jets are bent due to ram pressure stripping in a rich environment. 
\subsubsection{J125419.40+304803.0: Active}
Figure \subref{J125422.44+304} shows a clear central core, with
diffuse emission at 150 MHz surrounding the compact source. The
extended 150 MHz emission, on a much larger scale, shows an FR-I type
morphology, although relatively relaxed in shape. With the number of optical sources surrounding the radio source based on the  SDSS image, it is
possible therefore that this source is in a dense environment -- typically where FR-I sources are found at low redshift \citep{worr94}.
\subsubsection{J125931.85+333654.2: Active}
A faint core detected at $3\sigma$ seen in Figure \subref{J125930.81+333} indicates that the source is still active. The extended lobe emission as shown by the 150 MHz contours is much fainter for the eastern lobe than the western lobe, possibly indicating a large orientation angle of the radio source with respect to the plane of the sky.
\subsubsection{J130003.72+263652.1: Active}
A clear central core at 6 GHz is seen in Figure \subref{J130004.25+263}, with large extended emission at 150 MHz extending to the north and fainter emission in the south. The nature of the source (FR-I or FR-II) remains unclear due to the lack of high resolution observations at 1.4 GHz. The morphology is uncharacteristic of classical FR-I or FR-II sources, and this may suggest that this is a newly active source, with the 150 MHz contours in this image describing the ageing remnant plasma from previous activity.
\subsubsection{J130013.55+273548.7: Active (Misidentified)}
Figure \subref{J130014.65+273} does not show any compact emission at 6 GHz at the position of the optical ID. However, there is a compact source $\sim 8$ arcsec ($37$ kpc) to the south-west of the optical ID, which is more in line with the jet axis based on the positions of the radio lobes seen at 150 MHz. Assuming that this is the core, the optical ID for this source made by \citetalias{hard_lofar16} is incorrect. However, there is no currently detected optical ID at the position of the core in SDSS, possibly because the host galaxy is at a high redshift and/or dust obscured.
\subsubsection{J130332.38+312947.1: Active}
Figure \subref{J130332.47+312} shows a clear compact core at the position of the optical ID. The extended lobe structure at 150 MHz is that of a classical FR-II radio galaxy, and we therefore confirm this source as being active.
\subsubsection{J130415.46+225322.6: Active (Misidentified)}
Figure \subref{J130415.97+225} does not show any clear compact source
at 6 GHz close to the optical ID. There are however, two sources of
compact emission in the image -- adjacent to the northern lobe and a faint source $\sim 15$ arcsec above the southern lobe on the jet axis. It is likely that the latter is the radio core of this system and in this scenario therefore, this system is active. The proximity of this core to the southern lobe, relative to the northern lobe, may be explained if the southern lobe is oriented closer towards the line of sight than the plane of the sky, or if the source is intrinsically asymmetrical. The optical ID at the position of this 6 GHz compact source is J130414.25+225305.0, and we assign this as the new optical ID corresponding to this active radio source.     
\subsubsection{J130532.49+315639.0: Candidate remnant}
Figure \subref{J130532.02+315} shows no sign of a central core within the extent of the emission at 150 MHz, which seems to be diffuse and showing no signs of typical lobe structure seen in classical FR-I or FR-IIs. There are no obvious signs, based on the number of catalogued SDSS sources around the radio source, of a dense group or cluster environment. If this is a true remnant radio galaxy, it is likely that it has had a long remnant phase, based on the morphology shown in Figure \subref{J130532.02+315} alone.
\subsubsection{J130548.65+344052.7: Candidate remnant}
Figure \subref{J130548.81+344} shows no clear sign of a compact core, indicating that the core and jets have switched off. The 150 MHz contours show double-lobed structure extending to the east and west, although relaxed in shape which is typically expected of remnants.
\subsubsection{J130640.99+233824.8: Active}
Figure \subref{J130641.12+233} shows a clear compact source at 6 GHz at the position of the optical ID, with extended emission surrounding the core at 150 MHz and extending towards the north and south. The lobe emission seems to show a relaxed morphology, in contrast to the emission immediately surrounding the core. This may indicate a restarted source, although we do not detect hotspots.
\subsubsection{J130825.47+330508.2: Active (Misidentified)}
Figure \subref{J130826.28+330} shows no significant compact emission at the location of the optical ID. Compact hotspots at 6 GHz can be seen however, surrounded by bright double radio lobes at 150 MHz. A detailed inspection of the 6 GHz image reveals a very faint, yet significant ($3\sigma$), compact object equidistant from the two lobes and along the jet axis. Furthermore, the object lies within the faint extension of the base of the northern lobe (see Figure \subref{J130826.28+330}) towards the southern lobe. This object however, is not currently associated with an optical ID from the SDSS DR12 catalogue. A cross-check with the Two-Micron All Sky Survey \citep[2MASS;][]{2mass} and the Wide-field Infrared Survey Explorer \citep[WISE;][]{wise} catalogues also shows no significantly detected sources. It is likely therefore that the true optical host is a high redshift galaxy. We therefore dismiss the current optical ID for this active radio source.
\subsubsection{J130849.22+252841.8: Active}
Figure \subref{J130849.75+252} shows a faint 6 GHz core, surrounded by diffuse emission at 150 MHz. The morphology of the emission is relaxed in shape, and is typical of what is expected of aged remnant plasma. Given that a core is detected, however, we confirm that this source is still in its active phase.
\subsubsection{J130916.85+305118.3: Candidate remnant}
Figure \subref{J130916.02+305} shows no apparent core near the optical ID, although the 150 MHz contours show characteristics of a bright, double-lobed radio galaxy. However, inspection of the LOFAR 150 MHz image on a larger scale reveals a further pair of lobes to the eastern side of Figure \subref{J130916.02+305} with a similar morphology. It is possible that this source is therefore a DDRG, and the non-detection of a compact core within the jet axis of both pairs of lobes makes it possible that this source is a remnant DDRG. Nevertheless,  we confirm the remnant candidate status for this source.
\subsubsection{J130915.61+230309.5: Active}
Figure \subref{J130916.66+230} shows a clear, compact and bright core, surrounded by emission at 150 MHz. There is also an extension towards the north-west, which is likely to be a radio lobe.
\subsubsection{J130917.63+333028.4: Removed from sample}\label{J130917.74+333_removed}
Figure \subref{J130917.74+333} shows compact bright objects at 6 GHz associated with the lobe structures at 150 MHz, particularly in the northern lobe of this source. However, no compact object is visible in this image near the optical ID or along the assumed jet axis. There is however, an SDSS source at the precise location of the compact object seen at 6 GHz in the northern lobe (J130917.11+333049.8). This however, is classified spectroscopically as a quasi-stellar object (QSO), or quasar. Moreover, the 6 GHz emission associated with the southern lobe has two faint components, neither of which have an SDSS optical ID. It is plausible that the northern object is a distinct radio-loud quasar, while the southern source represents a high-redshift radio galaxy. Given that both sources individually do not meet our original AGN sample criteria, we therefore removed this source from our parent radio-loud AGN sample, as well as from our original candidate remnant sample.
\subsubsection{J131040.25+322044.1: Candidate remnant}
Figure \subref{J131040.03+322} shows no visible compact core at 6 GHz, and the 150 MHz contours show a relaxed morphology, with no clear separation of the radio emission from the lobes and the central host. Given the large physical extent of the source ($\sim 500$) kpc and its morphology, it is most plausible that this is an extremely old source in a long remnant phase.
\subsubsection{J131039.92+265111.9: Candidate remnant}
Figure \subref{J131040.52+265} shows no evidence of a central  compact core at 6 GHz, with the extended emission at 150 MHz likely corresponding to  lobed remnant emission. A closer inspection of the 6 GHz map reveals some significant compact emission associated with the radio emission at 150 MHz to the south-west of the target source. Furthermore, there are FIRST and NVSS detections associated with this emission. Since, after inspecting the 150 MHz image, there is also an eastern lobe associated with this background source, this is likely a background radio galaxy with a western hotspot. However, we do not detect any significant compact core emission at 6 GHz along the jet axis of this source, nor do we find FIRST or NVSS core detections. Although this source meets the physical selection criteria of our parent AGN sample, we are unable to constrain its optical ID due to the large number of SDSS-detected sources along the jet axis. We therefore interpret this source as a detection of another remnant radio galaxy, although the lack of an optical ID means that we cannot include it in our sample. Nevertheless, we can confirm the candidate remnant status of the original target source.
\subsubsection{J131235.35+331348.6: Active (Misidentified)}
Figure \subref{J131236.14+331} shows no clear 6 GHz emission from the optical ID, with some faint 150 MHz emission extending towards the north-west and south-east. There is however, a compact source at 6 GHz seen towards the north-west (see Figure \subref{J131236.14+331}), as well as an extension of the 150 MHz emission towards the north-west that extends beyond the image. There is an optical ID corresponding to this compact source (J131231.68+331436.3), implying that this is an active radio galaxy with this new optical ID. Figure \subref{newopticalidJ131236.14+331} displays the position of the new optical ID at the position of this 6 GHz core, with respect to the previous optical ID in Figure \subref{J131236.14+331}. 
\subsubsection{J131405.16+243234.1: Active (Misidentified)}
Figure \subref{J131405.90+243} shows a clear core at 6 GHz, surrounded by extended emission at 150 MHz. This source shows the morphology of a classical double radio galaxy, although the 150 MHz contours visually show some sign of outer radio lobes being part of a previous remnant phase. Faint emission associated with the inner lobes can be seen in an un-smoothed 6 GHz map, while 1.4 GHz FIRST detections are also present at these positions, supporting the double-double nature of this source. A faint extension of the outer eastern lobe towards the south can also be seen. We can also dismiss the current optical ID and confirm a new optical ID for this source as J131405.99+243240.1, cross-matching the position of the 6 GHz core and this new optical ID. Due to the proximity with the previous optical ID, the influence of a potential merger event therefore could have disrupted the previous AGN activity, with the subsequent gas infall due to the merger feeding the nucleus to produce the current activity.
\subsubsection{J131446.57+252819.8: Candidate remnant}
Figure \subref{J131446.83+252} shows no apparent 6 GHz core at the optical ID, although two hotspots in the east and west directions are seen, associated with the outer lobes at 150 MHz. Although a double-double nature for this source is plausible, a closer inspection of the 6 GHz un-smoothed map reveals a very faint compact object in line with the jet axis of the outer lobes. We therefore infer that this image contains two radio galaxies, with the source that has hotspots being active due to a very compact and faint core detected at 3$\sigma$ lying on the jet axis, and the other source with the inner lobe structure without a core being a remnant. However, there is no currently detected optical ID at the position of the 6 GHz compact object. Given the $3\sigma$ detection along the jet axis, and the fact that there is a small extension of the eastern lobe towards the proposed radio core, it is still likely that this faint object corresponds to the core of the active source. Since we cannot optically identify this new source, we do not include this in our sample. Nevertheless, we confirm the remnant status of the original source (J131446.83+252), since it still meets the original AGN sample criteria as detailed in Section \ref{intro}.
\subsubsection{J131536.30+310615.5: Removed from sample}
Figure \subref{J131537.33+310} shows no compact 6 GHz core or significant emission at 150 MHz resembling a radio galaxy. A lower resolution LOFAR map made by \citetalias{hard_lofar16} shows some faint diffuse emission that may resemble a radio galaxy, although it is probable that this is an imaging artefact. Given that the 2$\sigma$ contours of the higher resolution map shown in Figure \subref{J131537.33+310} do not bear any resemblance to typical jet or lobe structure, we are confident therefore that this is not a genuine AGN, and remove this source from the original parent radio-loud AGN sample.
\subsubsection{J131827.83+291658.5: Removed from sample}
Figure \subref{J131828.97+291} shows no significant 6 GHz emission
corresponding to the optical ID, although there is a faint object
towards the south west from the centre of the image (inspecting an
un-smoothed 6 GHz image). This compact source is in fact at the
position of an SDSS source -- namely J131827.29+291659.3. However, this
object is currently identified by SDSS as a star and the lack of
spectroscopic information available for this object means that it
cannot robustly be classified as a radio-loud quasar. Given the
brightness of the compact source at 6 GHz ($\sim 100$ mJy) and the
relative weakness of 150 MHz emission at this position, it is likely that 
this radio emission is associated with a flat-spectrum radio-loud
quasar. Nevertheless, we find that the 150 MHz emission seen in Figure
\subref{J131828.97+291} does not satisfy our flux criterion; the
source was included in the sample based on a lower resolution LOFAR image. We therefore remove this source from our sample, as explained in Section \ref{lofardata}.  
\subsubsection{J131833.81+291904.9: Removed from sample}
This source contains no significant compact emission near the optical ID at 6 GHz, as seen in Figure \subref{J131832.33+291}. There is some faint 6 GHz emission associated with the extended structure at 150 MHz, however this is likely to be due to  artefacts in the image from bright sources nearby. A closer inspection reveals that the 150 MHz emission is in fact an extension of  J131828.97+291 (Figure \subref{J131828.97+291}), and it is possible that two separate sources were seen by the lower resolution images made by \citetalias{hard_lofar16}. Nevertheless, similarly to J131828.97+291, this source does not satisfy our flux criterion and we therefore remove it from our sample, as explained in Section \ref{lofardata}.
\subsubsection{J132402.51+302830.1: Active}
Figure \subref{J132403.45+302} shows a very faint and compact source
at 6 GHz at the location of the optical ID, detected at the 3$\sigma$
level. This source is surrounded by extended and diffusive emission at
150 MHz, showing a morphology similar to a classical  FR-I radio
galaxy -- consistent with the low-redshift of the object (Table \ref{targetinfo}) and the large elliptical shape of the host galaxy as seen in the SDSS image. There is another compact region slightly north of the core that also has a significant detection with FIRST, but no corresponding optical ID and is therefore likely to be emission from the same source. Furthermore, the northern jet at 150 MHz seems to change direction and tail towards the west at the position of this compact object, suggesting that it is indeed associated with this radio galaxy. 
\subsubsection{J132602.06+314645.6: Candidate remnant}
Figure \subref{J132602.42+314} shows no signs of a compact core in the vicinity of the optical ID, and the 150 MHz emission shows a relaxed, yet double-lobed morphology. We therefore confirm the remnant candidate status of this source. 
\subsubsection{J132622.12+320512.1: Active}
Figure \subref{J132622.56+320} shows a clear compact core at 6 GHz centred on the optical ID. The 150 MHz extended emission displays a morphology typical of a classical FR-I radio galaxy. We therefore have robust evidence of the active nature of this source.
\subsubsection{J132738.77+350644.6: Active}
Figure \subref{J132735.32+350} shows this source has a classical FR-II
morphology, with a large extension at 150 MHz and hotspot emission
at 6 GHz (obscured by the contours in this image). Faint 6 GHz
emission can be seen near the optical ID, as seen in the zoom-in -- however, a positional cross-check with SDSS shows no sign of an optical host at this position. A closer inspection of the 6 GHz image reveals a brighter compact object along the jet axis towards the east of the current optical ID (unseen in this smoothed map due to obscuration by the 150 MHz contours). Cross-matching its position with SDSS reveals that there is in fact an optical ID associated with this object. Inspecting a colour-scale map of the SDSS image, the morphology of the host suggests that this may be a binary system, although this has not been confirmed by DR12. Nonetheless, we dismiss the current optical ID, and associate the radio source with the SDSS galaxy J132737.92+350650.2.
\subsubsection{J132949.25+335136.2: Active}
Figure \subref{J132948.87+335} shows very clear signs of a
prototypical FR-II radio galaxy -- a bright central core at 6 GHz and elongated bright lobes at 150 MHz. The radio core is at the location of the optical ID, while there are also 1.4 GHz FIRST and NVSS detections in the lobes. Given its morphology, FIRST and VLA core detections, 150 MHz flux density and its relatively high spectroscopic redshift of 0.5601 compared to the rest of the sources in our sample, this active radio galaxy represents one of the most powerful sources in our core-detected sample.
\subsubsection{J133016.12+315923.9: Candidate remnant}
Figure \subref{J133016.33+315} shows the lack of a significant detection of a compact source at 6 GHz at the optical ID, although the 150 MHz map shows a classical double lobe structure. The northern lobe shows an extension to the east, and the southern lobe shows an extension to the west.
\subsubsection{J133058.91+351658.9: Active}
Figure \subref{J133057.34+351} shows a clear, bright, compact core
with faint collimated extensions at 6 GHz. This is a known FR-II radio
galaxy  \citep{kozi11}. The lobe emission at 150 MHz   seems to be
dissipative with various extensions and asymmetric with the presumed
jet axis -- most notably of the southern lobe which extends directly
south and also has a separate extension towards the north west. This
extension is possibly a separate source -- there is a faint 6-GHz
detection within the 150 MHz contours at this position, coincident
with an SDSS optical ID (J133054.41+351657.8). The redshift of this galaxy ($z=0.489$) is higher than that of J133057.34+351 ($z=0.3158$), although the former is based on photometry and the latter on spectroscopy. It is possible therefore that Figure \subref{J133057.34+351} shows the image of two interacting radio galaxies. We suggest that this is a separate radio source previously unidentified by \citetalias{hard_lofar16}, but we find that the new radio source does not satisfy our flux density criteria ($>80$mJy), and we therefore do not include this as an additional source in our sample.  
\subsubsection{J133309.94+251045.2: Candidate remnant}
Figure \subref{J133310.56+251} shows two sources of radio emission at 150 MHz away from the central optical ID, likely corresponding to the two lobes of a radio galaxy. No 6 GHz emission can be seen in Figure \subref{J133310.56+251}; however, a closer inspection of the 6 GHz map shows some faint emission coinciding with the edge of the northern lobe at 150 MHz ($\sim 30$ arcsec from the central optical ID), and we interpret this as hotspot emission from the northern lobe. However, we do not see any clear evidence of a radio core at low- or high-frequencies in line with the jet axis, and therefore we can confidently confirm the remnant candidate status of this source, albeit with faint hotspot emission at high frequencies.  
\subsubsection{J133422.15+343640.5: Candidate remnant}
Figure \subref{J133422.21+343} shows what seem to be two very bright hotspots at 6 GHz, and associated bright lobes based on the 150 MHz contours. However, no core is apparent. It is possible that the presence of a nearby bright object to the south of this source is hindering the detection of a faint and compact core at the optical ID. After visually inspecting the 6 GHz map in detail, and considering the local RMS level at the optical ID is at the sensitivity limit of the sample ($\sim 10\mu$Jy), the core is clearly not detected. It is likely therefore that this 6 GHz snapshot image has captured the radio galaxy very soon after the core switched off, and that the hotspots are still detected at 6 GHz due to the last injection by the fading jets.
\subsubsection{J133502.36+323312.8: Active}
Figure \subref{J133502.38+323} shows bright, compact emission at 6 GHz at the location of the optical ID. Furthermore, the brighter emission towards the east agrees with the physical inference of the source being active, with an eastern hotspot. There are also 1.4 GHz FIRST detections at the hotspot region as well as at the western lobe, although the western hotspot is not detected in our 6 GHz map.
\subsubsection{J133642.53+352009.9: Active}
Figure \subref{J133643.09+352} shows a clear central compact object at 6 GHz at the location of the optical ID. There is also a 6 GHz detection to the west of the optical ID and well within the central area of the 150 MHz emission. A cross-check with SDSS reveals the latter is a radio-bright `star', and therefore we are confident of the current optical ID. Although this source has a core detection, the extended emission  shows a relaxed shape, similar to what would  be expected from remnants.
\subsubsection{J134702.03+310913.3: Active}
Figure \subref{J134701.71+310} shows a very faint, compact object at the centre of the image, coincident with the optical ID. Fitting an elliptical gaussian to this source, we detect the 6 GHz emission at the $\sim 5\sigma$ level, and confirm a core detection. The LOFAR 150 MHz contours show a relaxed morphology of the plasma ($\sim$ 400 kpc in extent), indicative of a remnant. However, since we do have a significant core detection at the centre of the radio source, we confirm its nuclear activity.  
\subsubsection{J134802.83+322938.3: Candidate remnant}
Figure \subref{J134802.70+322} shows no obvious core, or any other 6 GHz emission in the vicinity of the optical ID. The 150 MHz contours show clear extended structure on either side of the host galaxy, resembling a relaxed morphology and signs of the northern lobe diffusing. We therefore confidently confirm the remnant candidate status for this source. 
\section{Discussion}\label{discussion}
In this section we interpret our result on  the remnant fraction and its implications for the dynamics of radio-loud AGN and their duty cycles.
\subsection{Remnant fraction}\label{remnantfraction}
The results obtained in this paper, which aimed to develop a
systematic survey of radio-loud AGN remnants, have been obtained by
follow-up observations of low-frequency, wide-area observations of the
H-ATLAS field which initially provided a sample of candidate remnant
radio-loud AGN. To first order, these potential remnant sources were
identified by selecting those sources without detected cores in FIRST
images, giving 38 candidate remnants (remnant fraction of 38/127). The
resolution and sensitivity of FIRST made this remnant fraction an
upper limit, as discussed in Section \ref{intro}. Our new VLA
observations with sub-arcsec resolution at 6 GHz have enabled cores to
be detected in this candidate remnant sample (Figure \ref{jvlaimages}), giving
revised constraints on the AGN remnant fraction.

From our 34 remnant candidates (after removing four sources that did
not meet our sample criteria -- see Section \ref{lofardata}), we also remove the source J130917.74+333, shown in Figure \subref{J130917.74+333} (see Section \ref{J130917.74+333_removed}). Our final sample consists of 33 remnant candidates out of 122 radio-loud AGN detected with LOFAR. From these, we see no significant evidence for a radio core in 11. This puts the candidate remnant fraction at $11/122=9$ per cent, representing a significant decrease from the upper limit determined by \citetalias{hard_lofar16} and by \cite{brie17}. The implications of this result regarding the dynamics of remnant radio galaxies and the AGN duty cycle are given in Section \ref{dynamics}.

\subsection{Spectral indices}\label{specindices}

The radio spectrum of a radio galaxy gives  important
information on the radiative age of the electrons responsible for the
emission. For more accurate constraints on the AGN duty cycle, it is
therefore crucial to understand the evolution of the spectral shape
of radio galaxies in their remnant phase. In the absence of
nuclear activity and/or hotspots, remnant sources are expected to have low-frequency 
steep spectra, associated with the energy losses \citep{pach70}. However,
\cite{murg11,godf17,brie16,brie17} suggest that remnants as a population may show a wide
range of spectral characteristics, including flat low-frequency
spectral indices consistent with active sources, while also showing
steep spectra at higher frequencies. It is important to learn whether
these spectral features are characteristic of a significant fraction of remnants, or
simply isolated cases. 

\begin{figure}
\centering
\hspace*{-0.5cm}
\includegraphics[scale=0.5,trim={1cm 0 0 0},clip]{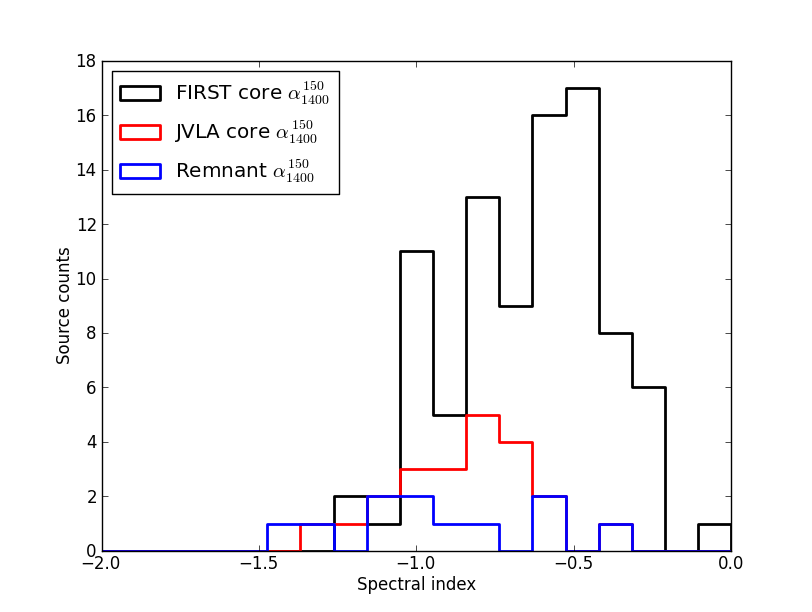}
\caption{Histogram of 150 MHz-1.4 GHz spectral indices for all sources in our sample colour-coded by core-detections. The candidate remnant with a flat spectral index $\alpha_{1400}^{150}=-0.4$ is only an upper limit due the lack of an NVSS detection at 1.4 GHz.}
\label{spixhist}
\end{figure}

A spectral analysis of our
remnant sample, comparing the 150 MHz-1.4 GHz integrated spectral indices of the
remnants (blue) to that of the active sources (red) in our sample,
is shown in Figure \ref{spixhist}. Spectral indices are based either on catalogued flux density values or on measurements directly from the LOFAR image at $150$~MHz, and the corresponding NVSS images at $1.4$~GHz. The $150$-MHz catalogue used was that from \citetalias{hard_lofar16} and a $1.4$-GHz source catalogue was generated by running \textsc{PyBDSF}\footnote{\url{http://www.astron.nl/citt/pybdsf}} \citep{pybdsf} on the NVSS image covering the same area as the LOFAR image. LOFAR/NVSS cutout images of each remnant candidate were then inspected visually in tandem; regions were defined for each source when the \textsc{PyBDSF} catalogued sources were judged to not accurately capture the source emission in that band. Typically LOFAR regions were identified for large sources where not all the flux was included in the catalogue, and NVSS regions were identified where the source was blended with a nearby source in the lower resolution NVSS image or was split into several sources in the NVSS catalogue. Total flux densities were extracted directly from the relevant images in the aperture defined by the region, and these aperture flux densities were used in preference to the catalogued values in determining the spectral indicies. LOFAR source sizes were determined by the largest extent of the $3\sigma$ threshold of each source. We also include in Figure \ref{spixhist} the parent AGN sample that have
FIRST-detected cores (black), to compare their spectral index
distribution with the active sources in our sample that have faint and
compact cores. We note that four sources (two core-detected and two candidate remnants) do not have an NVSS detection, and therefore have upper limits to their spectral indices only.

Based on the data presented in Figure \ref{spixhist}, the median spectral indices\footnote{Errors quoted are the standard errors of the median assuming a normal distribution.} are: $\alpha^{1400}_{150}\approx -0.97 \pm 0.03$ for the candidate remnants; $\alpha^{1400}_{150}\approx -0.80 \pm 0.01$ for the sources with VLA-detected cores; $\alpha^{1400}_{150}\approx -0.62 \pm 0.00$ for the sources with FIRST-detected cores, indicating a tendency for remnants, on average, to have a slightly steeper low-frequency spectral index than active sources. \cite{parm07} select their sample of remnants using $\alpha^{327}_{1400}< -1.3$ as a compromise to include the largest possible number of remnants while minimising the number of steep-spectrum active sources. However, it is clearly possible, based on the median spectral indices of the remnant candidates in our sample, for many remnants to be missed by these spectral selection methods, as suggested by \cite{murg11,brie16}. Moreover, it is possible for remnant samples to be contaminated with active sources that may possess steep spectra, based on this spectral index criterion alone. Figure \ref{spixhist} shows that the remnants and the active sources selected in this study clearly possess a wide range of spectral indices, and this is consistent with the conclusions of \cite{brie17}. We computed Wilcoxon-Mann-Whitney rank tests between the three samples to test whether  the remnants have significantly steeper spectral indices than the active sources. We found, at the 95 per cent confidence level, significantly steeper spectral indices for the candidate remnants  than the sources with FIRST-detected cores. We also found that the spectral indices of the sources with VLA-detected cores are significantly steeper than the sources with FIRST-detected cores.  The general trend that the sources with brigher, FIRST-detected cores tend to have flatter spectral indices than the sources with more compact cores may imply a positive correlation between core brightness and spectral index -- sources with brighter cores will tend to have extended emission less dominated by energy losses than those with faint and compact cores. However, we found no significant difference in steepness of spectral index between the candidate remnants and the sources with VLA-detected cores. Larger sample sizes will be needed in order to make these statistics robust -- our VLA core-detected  sample consists of only 22 sources compared to the FIRST core-detected sample of 122.

As a further check on our spectral index statistics, we find that five sources in our sample (remnant or active) have $\alpha^{150 }_{1400}\leqslant -1.2$, giving an ultra-steep-spectrum fraction of 4.1 per cent (5/122). This is comparable with the results of a LOFAR study of the Lockman Hole field \citep{maho16}, where an ultra-steep-spectrum fraction of 4.9 per cent was obtained. We note however, that this percentage includes all radio sources detected with LOFAR including radio-loud quasars and star-forming galaxies, and therefore acts as only an upper limit to the fraction of radio galaxies with ultra-steep-spectra. Nevertheless, these statistics demonstrate the problems with solely applying the classical ultra-steep spectrum criterion to select remnants, as found by \cite{brie16,brie17}, since we identify more than twice the number of remnant candidates (11) using our core-selection method than this spectral index selection method (5). A visual inspection of the LOFAR images of these five steep-spectrum sources shows that they have a range of morphologies, from relaxed to powerful FR-II-type, with only one being a candidate remnant \textit{and} having a relaxed morphology.
\subsection{Core prominence}\label{coreprom}
\begin{figure}
\centering
\hspace*{-0.5cm}
\includegraphics[scale=0.5,trim={1cm 0 0 0},clip]{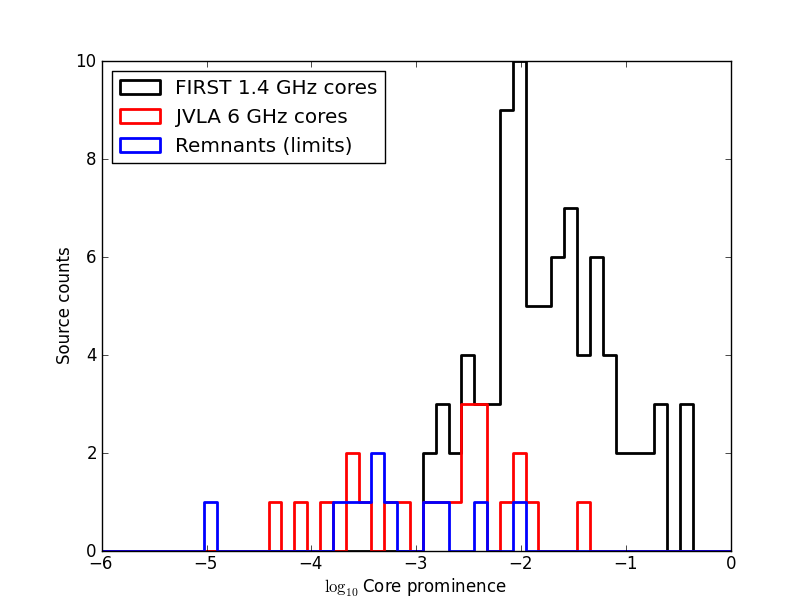}
\caption{Core prominences for FIRST-detected sources, VLA-detected sources and sources with non-detected cores (remnants), plotted in logarithmic scale. VLA core fluxes were obtained by fitting an elliptical gaussian to the corresponding 6 GHz radio maps. For non-detections (remnants) we use 3$\sigma$ upper limits on the flux at the positions of their current optical ID.}
\label{corepromhist}
\end{figure}

The core prominence, as explained in Section \ref{intro}, describes the brightness of a radio core relative to its extended
emission. Since we have sampled active sources with faint and compact
cores, we expect to find systematically lower core prominences than
those obtained with FIRST by \citetalias{hard_lofar16}. In Figure \ref{corepromhist} we
present the distribution of core prominences for our entire sample. To
determine the flux density of detected cores in our 6 GHz
observations, for each source, we fitted an elliptical Gaussian to a
small region in the map immediately surrounding the core. For the
total flux density of the extended emission of each source we use the
LOFAR 150 MHz flux density determined by \citetalias{hard_lofar16}. For
completeness, we also present 3$\sigma$ upper limits on the core
prominences of our 11 remnant candidates, defining upper limits on the
core flux density of the remnants based on the local RMS level within
a square box 20 pixels in size (dimensions much larger than the VLA beam size) at the location of the optical ID .

Figure \ref{corepromhist} shows that, as expected, the core prominences for our compact core-detected sample are significantly lower than the FIRST core-detected sample. The mean core prominence for the faint 6 GHz cores of $\sim 0.005$ is an order of magnitude lower than the average from the FIRST core detected sample, consistent with the idea that our sensitive, high resolution VLA observations have identified many active sources with faint and compact cores that would be missed by instruments with lower sensitivity and resolution. Since a few of the core prominences of our active sources are comparable to that of the upper limits on the candidate remnants, our results demonstrate that radio-loud AGN with bright, extended emission with faint cores can be misidentified as remnants using observations at lower resolution such as FIRST, if only core prominence is used as a selection criterion. 

We also derived core radio luminosities at 6 GHz, listed in Table \ref{targetphysicalparms}. In comparison to a sample of bright, nearby early-type galaxies formed by \cite{sadl89}, who infer that the core radio emission for most of their sample is indeed dominated by an active nucleus, the core powers of our sample are higher, including the upper limits on the candidate remnants. This is unsurprising, since our extended, higher-redshift sources are biased towards higher luminosities than that of \cite{sadl89}, and that some of our candidate remnants may in fact be active. Radio-loud AGN thus have a broad range of possible core luminosities in their active phase -- we cannot select a single value below which only remnant AGN may be found using our current sample.
\subsection{Power-linear size}\label{powersize}
\begin{figure}
\centering
\hspace*{-0.5cm}
\includegraphics[scale=0.40,trim={0.5cm 0 0 0},clip]{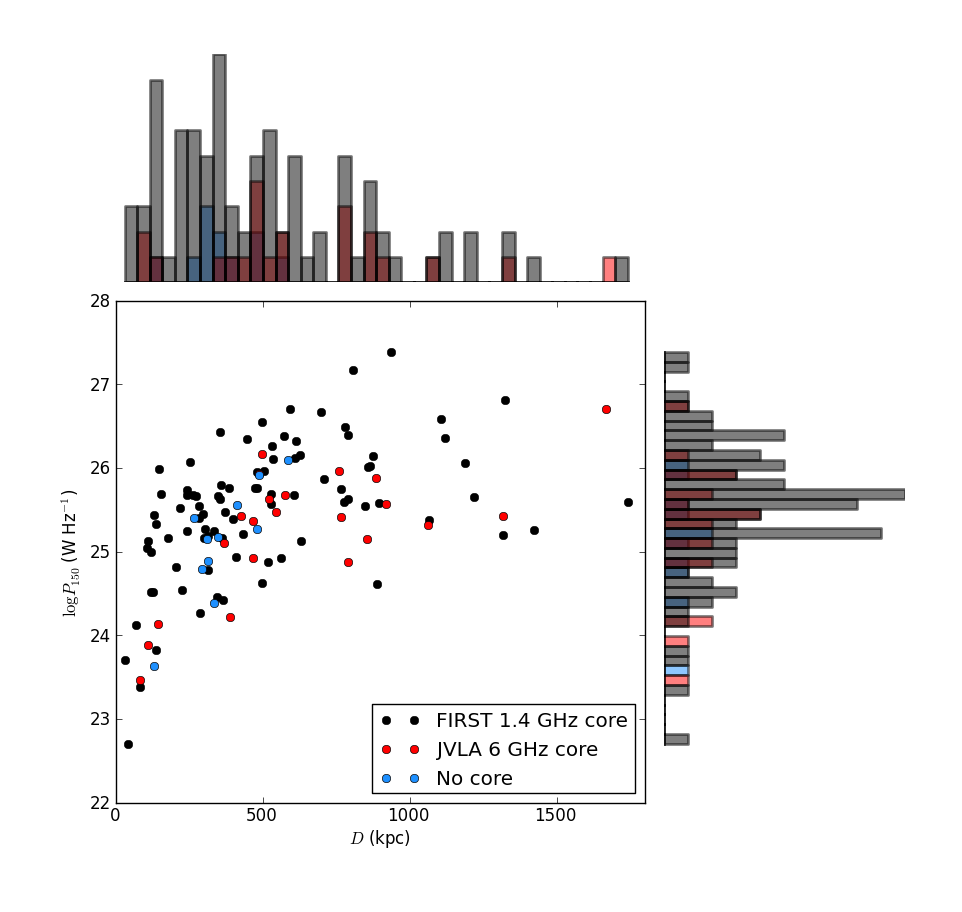}
\caption{Power-linear size diagram for all sources in this sample. The radio power $P_{150}$ is defined as the 150-MHz luminosity and the linear size $D$ defined as the largest physical extent of the radio source at 150 MHz. We also collapse the distribution of our sources to 1-dimensional histograms, for a clearer view of the distribution in radio power (vertical axis) and linear size (horizontal axis). Note that we have corrected the radio powers and sizes of the four optically misidentified sources (Table \ref{opticalmisidtable}).}
\label{powersizeplot}
\end{figure}
Figure \ref{powersizeplot} shows a power-linear size diagram  for the candidate remnant
sample (blue), 6 GHz core-detected sample (red) and the sources from
the parent AGN sample with FIRST-detected cores (black). To first
order, we see that our candidate remnants do not occupy a particularly
special position in power-size space with respect to the core-detected
sources, as also seen by \citetalias{hard_lofar16}. Performing a two-sample Kolmogorov-Smirnov (K-S) test, we find, at
a confidence level of $95$ per cent, that there is no
statistically significant difference in the distributions of radio
luminosities and linear sizes between the candidate remnants and the core-detected 
sources. We do however, find a significant difference in the physical sizes of
the candidate remnants and of the active sources with 6 GHz detected cores -- the
sources with 6 GHz cores tend to be larger than the candidate remnants for a
given radio power. This is surprising, since we might expect remnant
sources to be older, and therefore to have expanded to larger
distances than active sources, on average. Our analysis, on the other
hand, suggests otherwise. This implies that remnants do indeed fade
very rapidly, since the intrinsically larger remnants that have
experienced more energy losses would have escaped the flux criterion
imposed on our sample, and some may have also escaped detection by
LOFAR altogether. This would also explain the lack of significantly steeper spectral indices of the remnants compared to the sources with VLA-detected cores (Section \ref{specindices}), since the steepest-spectrum sources have escaped detection. Our results can be explained if the remnants in our sample
are relatively young compared to the rest of the population (captured
soon after switch-off), and therefore generally have physical sizes
and luminosities comparable to those of active sources. This is consistent with the conclusions of \cite{godf17} and of \cite{brie17} -- that the core prominence selection method would select young remnants  that would be otherwise missed by spectral selection methods. In this
scenario, we would expect to observe older and therefore larger
remnant sources, that would dominate samples selected by spectral index, in more sensitive surveys.

The radio power and linear size of a particular source are known to be degenerate  with other physical parameters, such as radio morphology and environment \citep{hakr13}. For example, it can be difficult observationally to distinguish between radio sources of a given power that are young and compact, and aged sources in rich environments whose expansion has been `confined' by a dense external medium \citep{an12,murg11}. This may also explain the lack of a substantial difference in power-size between remnants and active sources in Figure \ref{powersizeplot}. However, as can be shown by the histograms in Figure \ref{powersizeplot}, as well as the results of the K-S test, we see that all classes of sources in our sample are generally clustered around the same values of radio power and physical size. This might  suggest that our flux-limited remnant sample may be affected by a degeneracy between radio power/size and environment, although we cannot test the environmental dependency with existing observations. Future  observations with increased sensitivity, followed up by optical and/or X-ray surveys to determine the nature of their environment, may be useful in determining a clearer evolutionary path for remnants. 
\subsection{Optical misidentification}\label{opticalmisid}
As a sub-product of our VLA 6 GHz observations, we were able to
constrain and confirm the SDSS optical identifications of the sources
in this sample previously made by \citetalias{hard_lofar16}. By detecting
compact cores at high spatial resolution we have been able
cross-match their spatial position with the position of the current
optical ID, confirming their ID while also making new optical IDs for
sources with cores that correspond to different optical sources.

We have confirmed that sources J130415.97+225, J131405.90+243,
J130917.74+333 and J132735.32+350 were misidentified by
\citetalias{hard_lofar16}, and confirm their new optical IDs based on SDSS
DR12, detailed in Table \ref{opticalmisidtable}. This misidentification fraction of $\sim 10$ per cent is
high, but it should be borne in mind that these large sources without
any core detection in FIRST will be the most challenging of all to
optically identify in LOFAR surveys; most sources are unresolved or of
small angular size and for these the optical identifications are
robust. Nevertheless our work gives an idea of the magnitude of the
problem for the largest sources, in the absence of high resolution
observations that are sensitive to compact cores. Future wide-area
surveys such as LoTSS will need to consider this limit when optically
identifying radio-loud AGN that have low core prominences.

\subsection{Dynamics of remnant sources}\label{dynamics}
The relatively small upper limit to the  remnant fraction (9 per cent) imposed by our
observations has strong implications for the understanding of the AGN
duty cycle and the dynamics of radio-loud AGN. Previous work on understanding the radio-loud AGN duty cycle based on radiative modelling has suggested, in some cases, that the remnant phase may be comparable or even longer than the active phase \citep[e.g][]{parm07,brie16},
meaning that remnants should be at least as abundant as active sources
using sensitive instruments such as LOFAR. On the other hand, our
remnant fraction requires a very rapid fading process. A robust calculation of the duty cycle, of
course, would involve determining the spectral ages of all the remnant
sources in our sample which are expected to vary between $5\sim 100$
Myrs, an accurate analysis of which would require using modern, spatially resolved techniques that require deeper observations
at a broad range of frequencies with comparable resolution
\citep{harwood13} -- \cite{harw17} find that spectral ages derived from traditional integrated model fitting techniques give poor descriptions of a source's spectrum. Recent radiative age mapping of the remnant B2 0924+30 using spatially resolved techniques have shown that its active phase is around a factor of two longer than its remnant phase \citep{shul17}, agreeing with the general results of this paper. However, \cite{brie17} suggest that radiative ages of remnants would still only yield upper limits, given that the dynamical evolution of radio galaxies becomes an important factor in the remnant phase (also implied by the results of this paper) which is currently missing from models of spectral ageing. We suggest that adiabatic expansion losses must be incorporated into radiative models of radio source evolution, which may lead to more accurate constraints of the AGN duty cycle. Nevertheless, our results imply
that the remnant phase of radio-loud AGN is short, relative to the active phase. Therefore, the general picture we can describe for radio-loud AGN is that, once the jets switch off, the radio lobes are visible for a relatively short amount of time -- where \textit{both} radiative and adiabatic losses contribute to the shortening of the radiative lifetime. As the lobe plasma expands, its morphology may become relaxed, as seen by the LOFAR contours for most of the candidate remnants in Figure \ref{jvlaimages}. Physical processes within the host galaxy cause the jets to restart in a new active phase, quickly merging with the pre-existing remnant plasma.

Remnant radio-loud AGN hosted by field galaxies would be expected to fade
very quickly, whereas in richer and denser environments the lobes are
likely to retain the pressure balance with their environment over a
longer period, preventing rapid adiabatic losses \citep{murg11}. Given that the
physical sizes of the remnants in our sample are similar to those of active sources (Figure \ref{powersizeplot}), it is plausible that we are
only able to detect, even with the sensitivity of LOFAR, the youngest
and brightest remnants which may tend to reside in dense environments.
The effect of an environmental dependency on the analysis of our candidate remnants remains unclear without further observations. Nevertheless our results give a clear diagnosis for the lack of a
systematic sample of remnant sources until now, in that the
sensitivity of instruments and dynamic range are crucial factors in
the observations of remnant radio-loud AGN.

In our sample, we have clearly identified a candidate remnant without
a core possessing hotspots at 6 GHz --  Figure \subref{J133422.21+343} (J133422.21+343), similar to the candidate remnant 3C28 \citep{harw17}. This is unsurprising as hotspots are
expected to be bright for as long as the light travel time of the jet
once the source switches off, and supports the idea, presented in
Section \ref{powersize}, that the sensitivity limit imposed by current instruments
means that we are only able to capture remnants soon after switch-off.
We suggest that bright hotspots can be observed for the youngest remnants, although we reiterate that J133422.21+343 is still only a candidate remnant and may still have a compact core that is undetected by the VLA.

Table \ref{targetphysicalparms} lists the source properties of the 33 active and candidate remnant sources in our sample, including  their core flux density, core luminosity and spectral index of the low-frequency extended emission.

\section{Conclusions and future work}\label{conclusions}

In this paper we have provided a systematic survey of candidate remnant radio-loud AGN using the unprecedented capabilities of LOFAR and the VLA. Using the detection of compact cores, we are able to constrain the remnant fraction in the H-ATLAS NGP field obtained by \citetalias{hard_lofar16}. In summary we conclude that:
\begin{itemize}
\item The small candidate remnant fraction of radio-loud AGN (9 per cent) implies a relatively short fading time of the plasma left behind in the post-switch off phase, before the jets switch on again.
\item Our implied short fading time is in contrast to spectral modelling using continuous injection models, which derive longer remnant lifetimes than active lifetimes for a selection of dying sources. This may suggest that these ageing models must incorporate adiabatic losses \citep[as suggested by][]{harw17} for a more realistic description of remnants, confirming the results of the simulations made by \cite{brie17}.
\item Unbiased remnant selection strategies in future surveys (e.g. the LOFAR 150-MHz Tier 1 Survey) should bear in mind the uncertainties surrounding the spectral indices of the remnant population (i.e, recently switched off sources may still show flat spectral indices), while allowing the lack of a core detection as a strong proxy for remnant radio-loud AGN.
\item We suggest that our remnant sample may only contain the youngest of the population, as also implied by the results of \cite{godf17} and \cite{brie17}. We speculate that some sources  may reside in dense environments that prevent strong expansion losses, causing them to remain detectable with LOFAR.  
\end{itemize}
Future work may involve follow up observations of individual remnants from our sample to derive their spectral ages, in order to constrain the duty cycle for AGN. Environmental information is also desirable, with follow-up deep X-ray and/or optical observations potentially providing key constraints on the effect of environment on our sample.
\begin{table*}
\centering
\begin{tabular}{ccccccc}
\hline
Reference & SDSS Obj ID & SDSS name & \multicolumn{1}{p{1.5cm}}{\centering RA \\ (deg)} & \multicolumn{1}{p{1.5cm}}{\centering dec \\ (deg)} & r-band mag & \multicolumn{1}{p{1.5cm}}{\centering Redshift \\ z}\\
\hline \vspace{0.1mm} 
Figure \subref{J130415.97+225} & 1237667736123933636 & SDSSJ130414.25+225305.0 & 196.059 & 22.884 & 21.94 & 0.795 (p)\\
Figure \subref{J131236.14+331} & 1237665023836160370 & SDSSJ131231.68+331436.3 & 198.132 & 33.243 & 20.52 & 0.485 (s)\\
Figure \subref{J131405.90+243} & 1237667911672333044 & SDSSJ131405.99+243240.1 & 198.524 & 24.544 & 20.67 & 0.516 (s)\\
Figure \subref{J132735.32+350} & 1237664671644516542 & SDSSJ132737.92+350650.2 & 201.908 & 35.114 & 20.88 & 0.500 (p) \\
\hline
\end{tabular}
\caption{Table showing the properties and names of the new optical identifications made of four core-detected sources in our sample. We list only the properties of the new optical IDs, referencing the figures from Figure \ref{jvlaimages} in column 1, and for details of the previous optical ID we refer the reader to Table \ref{targetinfo} and Section \ref{remnantstatus}. Redshifts given are either photometric (p) or spectroscopic (s).}
\label{opticalmisidtable}
\end{table*} 
\begin{table*}
\centering
\begin{tabular}{cccccccc}
\hline
\rule{0pt}{1.0\normalbaselineskip}
LOFAR name & SDSS name & \multicolumn{1}{p{1.5cm}}{\centering 6 GHz core flux density \\ ($\mu$ Jy)} & \multicolumn{1}{p{1.5cm}}{\centering Flux density error \\ ($\mu$ Jy)} & \multicolumn{1}{p{1.5cm}}{\centering $\log$ 6 GHz core power \\ (WHz$^{-1}$)}  & $\alpha^{1400}_{150}$ & \multicolumn{1}{p{1.5cm}}{\centering $\log$ 150 MHz total power \\ (W Hz$^{-1}$)} & \multicolumn{1}{p{1.5cm}}{\centering Linear size \\ (kpc)}\\
\hline
J125143.00+332020.2 & SDSSJ125142.91+332020.6 & 442.82 & 21.083 & 23.292 & -0.9994 & 25.571 & 921.01\\
J125147.02+314046.0 & SDSSJ125147.03+314047.6 & 462.31 & 39.227 & 23.237 & -0.7919 & 25.633 & 522.53\\
J125311.08+304029.2 & SDSSJ125311.62+304017.3 & 19.958 & 1.2680 & 21.850 & -0.6406 & 24.925 & 467.17\\
J125419.40+304803.0 & SDSSJ125422.44+304728.1 & 59.711 & 37.156 & 21.773 & -0.7765 & 25.322 & 1063.1\\
J125931.85+333654.2 & SDSSJ125930.81+333646.9 & 209.23 & 18.536 & 23.284 & -1.0137 & 25.414 & 765.71\\
J130003.72+263652.1 & SDSSJ130004.25+263652.7 & 201.85 & 18.201 & 22.760 & -0.9195 & 25.362 & 469.07\\
J130013.55+273548.7 & SDSSJ130014.65+273600.0 & 184.82 & 22.325 & 22.740 & -0.9937 & 25.153 & 855.75\\
J130332.38+312947.1 & SDSSJ130332.47+312949.5 & 148.87 & 8.6801 & 23.450 & -0.6124 & 25.673 & 575.96\\
J130415.46+225322.6 & SDSSJ130414.25+225305.0 & 6.4460 & 3.800 & 22.155 & -0.7534 & 25.583 & 342.52\\
J130532.49+315639.0 & SDSSJ130532.02+315634.8 & <26.245 & & <21.918 & -0.9945 & 25.179 & 349.11\\
J130548.65+344052.7 & SDSSJ130548.81+344053.8 & <28.477 & & <21.614 & -0.5835 & 24.893 & 314.38\\
J130640.99+233824.8 & SDSSJ130641.12+233823.5 & 300.95 & 9.3260 & 22.416 & -0.5683 & 24.214 & 387.76\\
J130825.47+330508.2 & SDSSJ130826.28+330515.0 & 6.8182 & 6.7591 & 21.582 & -0.9181 & 25.422 & 427.12\\
J130916.85+305118.3 & SDSSJ130916.02+305121.9 & <20.957 & & <21.860 & -1.3005 & 25.269 & 480.09\\
J130915.61+230309.5 & SDSSJ130916.66+230311.3 & 525.18 & 2.2780 & 22.864 & -0.3507 & 24.136 & 144.66\\
J131040.25+322044.1 & SDSSJ131040.03+322047.6 & <61.650 & & <22.780 & -1.4250 & 26.092 & 584.94\\
J131039.92+265111.9 & SDSSJ131040.52+265105.5 & <25.706 & & <21.360 & <-0.3644 & 23.637 & 659.07\\
J131235.35+331348.6 & SDSSJ131231.68+331436.3 & 671.47 & 25.829 & 23.692 & -1.0759 & 25.762 & 1634.8\\
J131405.16+243234.1 & SDSSJ131405.99+243240.1 & 645.52 & 24.911 & 23.735 & -0.7281 & 25.715 & 637.17\\
J131446.57+252819.8 & SDSSJ131446.83+252820.5 & <34.176 & & <22.510 & -1.1086 & 25.403 & 267.44\\
J132402.51+302830.1 & SDSSJ132403.45+302822.8 & 74.599 & 11.863 & 20.612 & -1.2779 & 23.468 & 83.355\\
J132602.06+314645.6 & SDSSJ132602.42+314650.4 & <186.50 & & <22.451 & <-0.9743 & 24.387 & 333.47\\
J132622.12+320512.1 & SDSSJ132622.56+320502.3 & 335.72 & 5.9261 & 23.073 & <-1.1796 & 24.871 & 792.12\\
J132738.77+350644.6 & SDSSJ132737.92+350650.2 & 154.72 & 76.843 & 23.085 & -0.8113 & 26.709 & 1668.8\\
J132949.25+335136.2 & SDSSJ132948.87+335152.2 & 978.84 & 37.237 & 23.995 & -0.8557 & 25.883 & 884.27\\
J133016.12+315923.9 & SDSSJ133016.33+315919.7 & <317.95 & & <22.939 & -1.0848 & 25.552 & 412.15\\
J133058.91+351658.9 & SDSSJ133057.34+351650.2 & 124.15 & 20.506 & 22.546 & -0.6594 & 25.478 & 544.62\\
J133309.94+251045.2 & SDSSJ133310.56+251044.1 & <1.0794 & & <20.315 & -0.9218 & 25.148 & 311.21\\
J133422.15+343640.5 & SDSSJ133422.21+343634.8 & <25.727 & & <22.410 & -0.8020 & 25.910 & 489.14\\
J133502.36+323312.8 & SDSSJ133502.38+323313.9 & 70.684 & 26.627 & 22.594 & -0.7828 & 25.923 & 489.38\\
J133642.53+352009.9 & SDSSJ133643.09+352011.7 & 127.35 & 17.930 & 21.612 & <-1.0860 & 23.883 & 111.54\\
J134702.03+310913.3 & SDSSJ134701.71+310924.2 & 41.891 & 12.639 & 21.641 & -0.6901 & 25.105 & 370.29\\
J134802.83+322938.3 & SDSSJ134802.70+322940.1 & <30.713 & & <21.554 & -0.6274 & 24.787 & 293.52\\
\hline 
\end{tabular}
\caption{Table of source properties used for the analysis in Section \ref{discussion}. We also include the SDSS names of the optical host, updating the names for sources with new optical IDs. Candidate remnants only have upper limits to the 6 GHz core flux density, and do not have flux density errors. Core luminosities at 6 GHz are also given, based on the flux densities in column 3, redshifts in Table \ref{targetinfo} and Table \ref{opticalmisidtable}, and assuming a power-law spectral index of $-0.5$. $\alpha^{1400}_{150}$ gives the spectral index of the extended emission seen by LOFAR and NVSS. Sources that have no corresponding detection at 1.4 GHz with NVSS have spectral indices with upper limits only. $\log$ 150 MHz total power denotes the radio luminosity of the source based on the LOFAR images.}
\label{targetphysicalparms}
\end{table*}
\section*{Acknowledgements}
VM thanks the University of Hertfordshire for a research studentship
[ST/N504105/1]. MJH and WLW acknowledge support from the UK Science
and Technology Facilities Council [ST/M001008/1]. The research leading to these results has received funding from the European Research Council under the European Union's Seventh Framework Programme (FP/2007-2013) / ERC Advanced Grant RADIOLIFE-320745. TS acknowledges support from the ERC Advanced Investigator programme NewClusters 321271.

This work has made use of the University of Hertfordshire
high-performance computing facility
(\url{http://stri-cluster.herts.ac.uk}) and the LOFAR-UK computing
facility located at the University of Hertfordshire and supported by
STFC [ST/P000096/1]. We have made use of \textsc{astropy} \citep{astropy}, a core PYTHON
package for astronomy hosted at \url{http://www.astropy.org/}, and of
APLPY, an astronomical PYTHON package for plotting and visualisation
hosted at \url{http://aplpy.github.com}.

The National Radio Astronomy Observatory is a facility of the National Science Foundation operated under cooperative agreement by Associated Universities, Inc.





\label{Bibliography}

\lhead{\emph{Bibliography}} 

\bibliographystyle{styles/mnras}

\bibliography{references} 




\bsp	
\label{lastpage}
\end{document}